\documentclass[12pt,preprint]{aastex}

\usepackage{natbib}
\bibpunct[]{(}{)}{,}{a}{}{,}

\usepackage{amsmath}
\usepackage{amsfonts}
\usepackage{amsthm}

\newcommand{\dif}{\mathrm{d}}
\usepackage{bm}
\renewcommand{\vec}[1]{\bm{#1}}
\DeclareMathAlphabet{\mathsfsl}{OT1}{cmss}{m}{sl}

\DeclareMathOperator{\expn}{Mean}
\DeclareMathOperator{\medn}{Median}

\DeclareMathOperator{\ndist}{N}
\DeclareMathOperator{\betadist}{B}
\DeclareMathOperator{\expdist}{E}
\DeclareMathOperator{\gamdist}{G}

\DeclareMathOperator{\bindist}{Bin}

\DeclareMathOperator{\bindens}{Bin}
\DeclareMathOperator{\betadens}{B}

\newcommand{\nd}{\ensuremath{n_\mathrm{d}}}
\newcommand{\nc}{\ensuremath{n_\mathrm{c}}}
\newcommand{\eya}[3]{\ensuremath{\hat{y}_{\mathrm{#1},#2}{(#3)}}}
\newcommand{\ya}[3]{\ensuremath{y_{\mathrm{#1},#2}{(#3)}}}
\newcommand{\ey}[2]{\ensuremath{\hat{y}_{\mathrm{#1},#2}}}
\newcommand{\y}[2]{\ensuremath{y_{\mathrm{#1},#2}}}

\newcommand{\p}[2]{\ensuremath{\pi_{#1}{(#2)}}}
\newcommand{\pmin}{\ensuremath{p_{\min}}}
\newcommand{\muminp}{\ensuremath{\hat{\mu}_{\min p}}}

\newcommand{\ceil}[1]{\ensuremath{\lceil #1 \rceil}}
\newcommand{\floor}[1]{\ensuremath{\lfloor #1 \rfloor}}

\begin{document}
\bibliographystyle{toby}
 
\title{Bayesian Method for Disease QTL Detection and Mapping, using a
  Case and Control Design and DNA Pooling}

\author{Toby Johnson\altaffilmark{1}}
\affil{School of Biological Sciences, The University of Edinburgh}
\affil{West Mains Road, Edinburgh, EH9 3JT}
\email{toby.johnson@ed.ac.uk}

\altaffiltext{1}{Jointly affiliated to Rothamsted Research and to The University of Edinburgh}

\slugcomment{draft \today}

\begin{abstract}
  This paper describes a Bayesian statistical method for determining
  the genetic basis of a complex genetic trait.  The method uses a
  sample of unrelated individuals classified into two groups, for
  example cases and controls.  Each group is assumed to have been
  genotyped at a battery of marker loci using a laboratory effort
  efficient technique called DNA pooling.  The aim is to detect and
  map a quantitative trait locus (QTL) that is \emph{not} one of the
  typed markers.  The method works by conducting an exact Bayesian
  analysis under a number of simplifying population genetic
  assumptions that are somewhat unrealistic.  Despite this, the method
  is shown to perform acceptably on datasets simulated under a more
  realistic model, and furthermore is shown to outperform classical
  single point methods.
\end{abstract}

\section{Introduction}
For many traits of interest, including susceptibility to many genetic
diseases, the genetic basis is complex, meaning that many genes of
individually small effect (quantitative trait loci; QTLs) contribute.
For detecting and mapping such QTLs, association mapping studies that
use large samples of essentially unrelated individuals may have two
advantages over linkage studies that use pedigrees or families: For a
given sample size, association studies may have more power to detect a
QTL \citep[e.g.][]{rischmerikangas1996,risch2000}, and they may allow
the QTL to be fine mapped with greater resolution or precision
\citep[e.g.][]{terwilliger1995,mcpeek1999}.  One important
experimental design is a genome wide scan, in which many thousands of
markers covering the whole genome are typed \citep[see
e.g.][]{lander1996,rischmerikangas1996,kruglyak1999,carlson2003}.  The
aim may be to detect QTL that are not candidate genes and that have
escaped detection in linkage studies.  After preliminary analysis,
attention may focus on relatively small chromosomal regions, each
containing perhaps only one QTL.  Because of
nongenetic factors and the effects of other genetically distant QTLs,
each focal QTL will explain only a small fraction of the variance in
phenotype.  If the trait is binary (such as presence or absence of a
disease), the difference in QTL allele frequencies between the two
trait groups will be small.  The large numbers of markers required for
a genome wide scan will need to have been typed in large numbers of
individuals to detect such a QTL, and to infer its position,
frequency, and effect on the trait.

In order to reduce the cost of such a study, an experimental strategy
called DNA pooling has been proposed
\citep[e.g][]{arnheim1985,barcellos1997,sham2002,norton2004}.  Here
DNA from individuals with similar phenotypes is physically mixed
together into a pool before genotyping.  After overheads to do with
construction of pools and assay development, the cost of genotyping an
entire pool at a given marker is reduced to the cost of genotyping a
single individual.  Thus costs may be reduced by a factor close to the
number of individuals in a pool (divided by the number of experimental
replicates used for each pool), when the overheads can be spread
across many markers and across many disease studies.

In this paper I consider the simplest experimental design where there
are two trait groups.  These could have been classified by the
presence or absence of binary trait such as a disease.  Alternatively,
if the trait is quantitative, individuals from each tail of the trait
distribution could make up the two groups, for example the upper and
lower 10\% tails \citep[e.g.][]{darvasisoller1994,bader2001}.  For
simplicity I will refer throughout the paper to the two trait groups
as cases and controls, and to one of the alleles at the QTL as the
disease allele.  I note that data collected from more than two groups
can be analysed using the method developed here, by discarding or
combining data from some of the groups.  (An extreme example is where
each pool contains two chromosomes, i.e.\ genotype data are
available.)  Such an approach is valid and may be useful, but will
almost certainly not make most efficient use of the available data,
since it will be based on only a marginal observation that is almost
certainly not sufficient.

Compared with individual genotyping, a DNA pooling strategy incurs
three types of information loss and error.  At each marker only the
marginal counts of the alleles present within each pool are are
available, and so there is (i) no information about deviations from
Hardy--Weinberg proportions within each marker within each pool
\citep{rischteng1998} and (ii) no information about phase or linkage
information across markers within each pool \citep{johnson2005a}.
Further (iii), the marginal counts are only estimated from some kind
of quantitative genotyping experiment \citep[e.g.][]{germer2000,lehellard2002},
rather than by counting individual genotyping experiments.

The present approach deals with (iii) by allowing a quite general
class of models for errors in allele frequency estimation.  It
attempts to deal with (i) and (ii) by using the full likelihood given
the available data.  This likelihood does however assume a model that
makes simplifying assumptions that are not as realistic as one would
like.  This may be an acceptable price to pay, because it allows all
the necessary computations to be done analytically or using simple
numerical algorithms.  The aim is to develop a method that can be used
on very large data sets, with hundreds of cases and hundreds of
controls typed at hundreds of markers.

The simple model used here is a special case of the model introduced
by \citet{mcpeek1999} and studied by \citet{morris2000},
\citet{zhangzhao2000,zhangzhao2002} and \citet{liu2001} for haplotype
and genotype data.  In brief summary, I assume a unique mutation
event (perhaps a single nucleotide polymorphism, or a deletion)
generated the disease allele at the disease QTL, that there is a star
shaped genealogy since that mutation, that the disease allele is
absent from the control group, and that Hardy--Weinberg and linkage
equilibrium apply in the control group.

Avoiding the use of more complicated algorithms such as Markov chain
Monte Carlo (MCMC) means that there are no concerns about mixing and
convergence, and no difficulties with computing normalising constants
and Bayes factors for model testing and model comparison.  When the
model is correct, the Bayes factor is \citep[in various classical
senses, see][ ch.5]{ohaganforster} an optimal test statistic for
detecting a QTL \citep[see also][]{patterson2004}.  Therefore the
present approach can be viewed as choosing an approximate model that
is simple enough to allow an exact and optimal analysis.  It will
therefore complement alternative approaches that perform approximate
or sub-optimal analyses for more realistic models.

One example of such an alternative approach is to perform a classical
single point analysis, which applies a separate test to the data for
each marker.  Such an approach can be made essentially free of any
population genetic model, meaning that no worring assumptions are made
but also that there is no efficient framework for combining analyses
across many markers (e.g.\ to correct for multiple testing).  When
genotype data are available and when the QTL is \emph{not} one of the
typed markers, such single point methods are known to lack power
\citep[e.g.][]{risch2000,mott2000,zollnerpritchard2005}, and may
produce inefficient point estimates and inefficiently large region
estimates for the position of the QTL
\citep[e.g.][]{morris2002,morris2003}.

One justification for the method developed here is that although the
model is simple, it is to my knowledge the most complex model for
which a Bayesian analysis has been implemented using data from DNA
pools.  To provide further justification and reassurance about the
simplifying assumptions made, I have tested the method on data
simulated under a more realistic full coalescent model.  The
(necessarily classical) measures of performance are encouraging in
three respects.  Firstly, the power to detect a QTL is substantially
higher than for classical single point analysis.  Secondly, point
estimates for the position of the QTL derived from the present method
outperform the simple procedure of choosing the map position of the
marker with the most significant single point test result \citep[the
``minimum $p$-value method'',
e.g.][]{kaplanmorris2001a,kaplanmorris2001b}.  Thirdly, after
``flattening'' the posterior using a factor \citep[derived
by][]{mcpeek1999} that corrects for the fact that the true genealogy
is not in fact star shaped, credibility intervals for the position of
the QTL cover its true position with frequency equal to their nominal
size.  That is, they are well calibrated.

The structure of this paper is as follows.  In
section~\ref{sec:appr-model-prior} I review the model introduced by
\citet{mcpeek1999}, for which an exact Bayesian analysis can be
performed using multilocus estimated allele frequency data, collected
using DNA pools.  In section~\ref{sec:analysis} I describe in detail
how the computations for such an analysis are performed.  In
section~\ref{sec:simul-model} I review a more realistic coalescent
model that I have used to generate simulated data sets on which to
test the current approach.  In section~\ref{sec:results} I describe
the performance of the method developed in
sections~\ref{sec:appr-model-prior}--\ref{sec:analysis} on those
simulated data sets.  In section~\ref{sec:discussion} I summarise the
results and discuss how the method could be improved.

\section{The Simplified Model and Prior}
\label{sec:appr-model-prior}
\begin{figure}[p]
  \begin{center}
    \begin{picture}(0,0)%
      \includegraphics{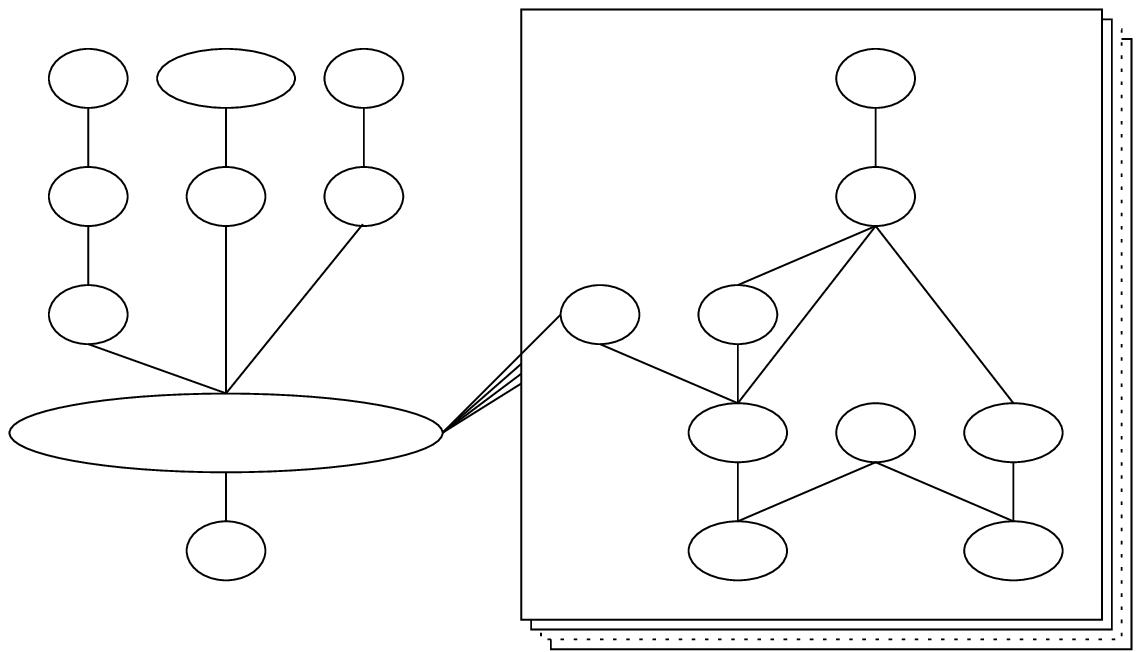}%
    \end{picture}%
    \setlength{\unitlength}{4144sp}%
    \begingroup\makeatletter\ifx\SetFigFont\undefined%
    \gdef\SetFigFont#1#2#3#4#5{%
      \reset@font\fontsize{#1}{#2pt}%
      \fontfamily{#3}\fontseries{#4}\fontshape{#5}%
      \selectfont}%
    \fi\endgroup%
    \begin{picture}(5697,3186)(-3164,-1918)
      \put(-2429,-961){\makebox(0,0)[lb]{\smash{{\SetFigFont{12}{14.4}{\familydefault}{\mddefault}{\updefault}{$\vec{x}=(x_1,x_2,\ldots,x_L)$}%
            }}}}
      \put(-2294,119){\makebox(0,0)[lb]{\smash{{\SetFigFont{12}{14.4}{\familydefault}{\mddefault}{\updefault}{$\rho$}%
            }}}}
      \put(-1664,119){\makebox(0,0)[lb]{\smash{{\SetFigFont{12}{14.4}{\familydefault}{\mddefault}{\updefault}{$\mu$}%
            }}}}
      \put(-1034,119){\makebox(0,0)[lb]{\smash{{\SetFigFont{12}{14.4}{\familydefault}{\mddefault}{\updefault}{$\tau$}%
            }}}}
      \put(-2339,-421){\makebox(0,0)[lb]{\smash{{\SetFigFont{12}{14.4}{\familydefault}{\mddefault}{\updefault}{$x_\mu$}%
            }}}}
      \put(-1799,659){\makebox(0,0)[lb]{\smash{{\SetFigFont{12}{14.4}{\familydefault}{\mddefault}{\updefault}{prior}%
            }}}}
      \put(1261,119){\makebox(0,0)[lb]{\smash{{\SetFigFont{12}{14.4}{\familydefault}{\mddefault}{\updefault}{$\pi_i$}%
            }}}}
      \put(1261,659){\makebox(0,0)[lb]{\smash{{\SetFigFont{12}{14.4}{\familydefault}{\mddefault}{\updefault}{$P_i$}%
            }}}}
      \put(631,-421){\makebox(0,0)[lb]{\smash{{\SetFigFont{12}{14.4}{\familydefault}{\mddefault}{\updefault}{$a_i$}%
            }}}}
      \put(  1,-421){\makebox(0,0)[lb]{\smash{{\SetFigFont{12}{14.4}{\familydefault}{\mddefault}{\updefault}{$x_i$}%
            }}}}
      \put(1261,-961){\makebox(0,0)[lb]{\smash{{\SetFigFont{12}{14.4}{\familydefault}{\mddefault}{\updefault}{$e_i$}%
            }}}}
      \put(-179,1109){\makebox(0,0)[lb]{\smash{{\SetFigFont{12}{14.4}{\familydefault}{\mddefault}{\updefault}{$i=1,2,\ldots,L$}%
            }}}}
      \put(586,-1501){\makebox(0,0)[lb]{\smash{{\SetFigFont{12}{14.4}{\familydefault}{\mddefault}{\updefault}{$\hat{y}_{\mathrm{d},i}$}%
            }}}}
      \put(586,-961){\makebox(0,0)[lb]{\smash{{\SetFigFont{12}{14.4}{\familydefault}{\mddefault}{\updefault}{$y_{\mathrm{d},i}$}%
            }}}}
      \put(1846,-961){\makebox(0,0)[lb]{\smash{{\SetFigFont{12}{14.4}{\familydefault}{\mddefault}{\updefault}{$y_{\mathrm{c},i}$}%
            }}}}
      \put(1846,-1501){\makebox(0,0)[lb]{\smash{{\SetFigFont{12}{14.4}{\familydefault}{\mddefault}{\updefault}{$\hat{y}_{\mathrm{c},i}$}%
            }}}}
      \put(-2339,659){\makebox(0,0)[lb]{\smash{{\SetFigFont{12}{14.4}{\familydefault}{\mddefault}{\updefault}{$R$}%
            }}}}
      \put(-1079,659){\makebox(0,0)[lb]{\smash{{\SetFigFont{12}{14.4}{\familydefault}{\mddefault}{\updefault}{$T$}%
            }}}}
      \put(-1709,-1501){\makebox(0,0)[lb]{\smash{{\SetFigFont{12}{14.4}{\familydefault}{\mddefault}{\updefault}{$\vec{m}$}%
            }}}}
      \put(-3149,659){\makebox(0,0)[lb]{\smash{{\SetFigFont{12}{14.4}{\familydefault}{\mddefault}{\updefault}{prior:}%
            }}}}
      \put(-3149,-1501){\makebox(0,0)[lb]{\smash{{\SetFigFont{12}{14.4}{\familydefault}{\mddefault}{\updefault}{data:}%
            }}}}
    \end{picture}%
  \end{center}
  \caption{Hierachical or Bayesian network structure of the model.  The region inside the
    rectangle is duplicated for each of $L$ SNPs.  Lines indicate the
    dependence structure of the model:  Variables not connected are
    independent, conditional on all other variables in the model.}
  \label{fig:factorisemodel}
\end{figure}

The main notations and abbreviations used are listed in
table~\ref{tab:notations-used}, and the conditional independencies of
the variables are represented in figure~\ref{fig:factorisemodel}.

I assume the data (collectively denoted $\hat{\vec{y}}$) consist of
allele frequency estimates at $L$ single nucleotide polymorphisms
(SNPs), obtained from a single pool of \nd{} case chromosomes and a
single pool of \nc{} control chromosomes.  Let $m_i$ be the known map
position of the $i$-th SNP.  Let the alleles at each SNP be
arbitrarily labelled 0 and 1, with $\eya{d}{i}{1}\equiv
\nd-\eya{d}{i}{0}$ the estimated count of the 1 allele at the $i$-th
SNP in the cases, and $\eya{c}{i}{1}\equiv \nc-\eya{c}{i}{0}$ the
estimated count of the 1 allele at the $i$-th SNP in the controls.
When there is no ambiguity, $\hat{y}$ will mean the estimated count of
the 1 allele at some SNP in some pool that is to be inferred from the
context.  (This can easily be generalised to let $\ey{d}{i}$ be
arbitrary vector valued information about the counts of the two
alleles at the $i$-th SNP in the cases, etc.)  The method may be
generalised to allow more than two alleles at each marker.

Let $\ya{d}{i}{1}$ and $\ya{c}{i}{1}$ be the true counts of the 1
allele at the $i$-th SNP in the cases and in the controls
respectively.  I assume that an error model
$\Pr{(\ey{d}{i},\ey{c}{i}|\y{d}{i},\y{c}{i})}$ has been chosen and
parameterised, for example from a calibration data set consisting of
pairs $(\hat{y},y)$ that were obtained from a given set of individuals
by genotyping them as a pool and by genotyping them individually.
Note that the error model cannot be factorised unless we assume that
the errors in the case and control pools are unconditionally
independent.  However, we may believe that the data were obtained
using assays that vary in precision across SNPs.  Letting $e_i$
indicate the unknown precision of the assay used for the $i$-th SNP,
we could model these beliefs as
\begin{equation}
  \Pr{(\ey{d}{i},\ey{c}{i}|\y{d}{i},\y{c}{i})}=\sum_{e_i}{\Pr{(\ey{d}{i}|\y{d}{i},e_i)}\Pr{(\ey{c}{i}|\y{c}{i},e_i)}\Pr{(e_i)}}\;\mbox{,}
  \label{eq:errorfactorise}
\end{equation}
that is, the error distribution is modelled as a mixture distribution
where each component of the mixture factorises.  In the following,
reasonable flexibility in the nature of this error model is allowed:
the errors must be independent across SNPs but they need not be
independent across pools and they do not need to be identically
distributed across SNPs.  A example error model is given in
section~\ref{sec:simul-model}.

Let the unknown map position of the disease QTL relative to the
leftmost marker SNP be $\mu$; often this will be the variable of main
interest.  I allow any prior $\Pr{(\mu)}$; obvious choices would be a
uniform density on physical distance, or one based on known gene
density \citep[see e.g.][]{rannalareeve2001}.  I assume a unique
mutation event generated the disease allele at the disease QTL, so
some number of haplotypes ($x_\mu$) in the case pool carry the disease
allele and are identical by descent (i.b.d.) at this position.  I
assume $x_\mu\sim\bindist{(\nd,\rho)}$ for a rate $\rho$,
and since $\rho$ itself is uncertain I assume a beta prior with
parameter $R=(R_1,R_0)$, so the prior mean is $R_1/(R_0+R_1)$.
Specifying $(R_1,R_0)=(1,1)$ specifies a flat prior on $[0,1]$ for
$\rho$.  I assume that the disease allele is absent from the control
pool.  The adequacy of this approximation, and ways in which it could
be relaxed, are taken up in the discussion in section~\ref{sec:discussion}.

Each disease allele is embedded in a block of i.b.d.\ haplotype, the
``ancestral haplotype'', with breakpoints at distances $d_\mathrm{L}$
and $d_\mathrm{R}$ to the left and right of the position of the QTL.
I assume a star shaped genealogy for the disease allele, so that
$d_\mathrm{L}$ and $d_\mathrm{R}$ for all blocks of ancestral
haplotype are conditionally independent and identically distributed
(i.i.d.) from an exponential distribution with mean $1/\tau$ Morgans,
where $\tau$ is the age of the disease allele in generations and
distances are measured in Morgans \citetext{\citealt{mcpeek1999}, see
  also \citealt{morris2000,zhangzhao2000,zhangzhao2002,liu2001}}.
(The left and right breakpoints are the positions of the nearest
crossovers in $\tau$ meioses where crossovers occur as a Poisson
process with rate 1.)  I allow any prior $\Pr{(\tau)}$; lognormal and
gamma are reasonable choices.  Specifying an exponential prior with
sufficiently small parameter $T$ (so the prior mean $1/T$ is
sufficiently large) specifies a prior that is effectively flat over
the region where the likelihood is non-negligible.  This has the
effect of making the posterior model probability or Bayes factor
proportional to $T$.  A crucial variable in the analysis is $x_i$, the
number of chromosomes in the case pool that carry the i.b.d.\
haplotype at the $i$-th SNP.  The assumptions above specify a
distribution for the $x_i$.  A key feature of the present method is
that the $x_i$ are not independent across SNPs, in constrast to what
is assumed in composite likelihood methods
\cite[e.g.][]{terwilliger1995,xiong1997,collins1998,maniatis2004,maniatis2005}.

All other non-i.b.d\ haplotype is called heterogenous ``non-ancestral
haplotype''.  The allele present in non-ancestral haplotype is assumed
conditionally independent across chromosomes within SNPs, and across
SNPs within chromosomes, with \p{i}{1} the probability of a 1 allele
at the $i$-th SNP and $\p{i}{0}\equiv 1-\p{i}{1}$.  (This is
equivalent to assuming Hardy--Weinberg and linkage equilibrium in
blocks of non-ancestral chromosome.)  This assumption is a very bad
one when haplotype or genotype data are available
\citep{liu2001,morris2002,listephens2003}, but when only multilocus
allele frequency data are available it may be more innocuous; this
assumption leads to a massive simplification in the likelihood.  I
assume an independent beta prior for each $\p{i}{1}$, with parameter
$P_i=(P_{i,1},P_{i,0})$, so the prior mean is
$P_{i,1}/(P_{i,0}+P_{i,1})$.  The method should be robust to
misspecification of $P_i$ as long as both elements are much smaller
than $\nc$.

The allele present on the ancestral haplotype at the position of the
$i$-th SNP, $a_i$, is assumed to be a single draw from the same
distribution as for a block of nonancestral chromosome.  That is, each
$a_i$ is an independent Bernoulli variable with parameter $\p{i}{1}$.

\section{Analysis}
\label{sec:analysis}
\subsection{Overview}
The purpose of the analysis is to compute the posterior distribution
for quantities of interest.  Here I focus on computing the Bayes
factor in favour of a model in which a QTL is present, versus a model
with no QTL.  This Bayes factor allows the posterior model
probabilities to be computed for any prior on the two models.  I also
compute the posteriors for $\mu$, $\tau$ and $\rho$, given that a QTL
is present, marginalising all other variables.

The model has a hierarchical structure as shown in
figure~\ref{fig:factorisemodel}.  Note that the joint distribution of
the $x_i$ depends on $\mu$, $\tau$ and $\rho$, but that the
probability of the data at the $i$-th SNP only depends on $x_i$ and
other variables $\pi_i$ and $a_i$ for which the prior is independent
across SNPs.  The first step of the analysis, in section
\ref{sec:at-each-snp}, is to perform an independent calculation for
each SNP that integrates out $\pi_i$ and $a_i$ conditional on $x_i$.
The second step of the analysis, in section
\ref{sec:hidden-markov-model}, is to integrate out all the $x_i$ and
$x_\mu$ conditional on $\mu$, $\tau$ and $\rho$.  This gives the
posterior density for these three variables.  The final step, in
section \ref{sec:final-posterior}, is to compute the marginal
posteriors for each of $\mu$, $\tau$ and $\rho$, and the normalising
constant or probability of the data given models with and without a
QTL.

\subsection{Calculations for the $i$-th SNP}
\label{sec:at-each-snp}
Note that all probabilities in this section are conditional on $x_i$,
the number of chromosomes in the case pool carrying the ancestral
i.b.d.\ chromosome.  At the $i$-th SNP let $\ya{d}{i}{1}\equiv
\nd-\ya{d}{i}{0}$ be the (unknown) actual count of the 1 allele in
cases, and $\ya{c}{i}{1}\equiv \nc-\ya{c}{i}{0}$ be the actual
estimated count of the 1 allele in controls.  Let
$\y{d}{i}=(\ya{d}{i}{1},\ya{d}{i}{0})$ and a similar notation apply
for controls.  Under the modelling assumptions we have
\begin{eqnarray}
  \label{eq:ydprob}
  \Pr{(\y{d}{i}|x_i,a_i,\pi_i)} &=&
  \bindens{(\ya{d}{i}{a_i}-x_i,\nd-x_i,\pi_i(a_i))} \\
  \label{eq:ycprob}
  \Pr{(\y{c}{i}|\pi_i)} &=&
  \bindens{(\ya{d}{i}{1},\nc,\pi_i(1))}
\end{eqnarray}
where $\bindens{(x,n,p)}$ is the probability of observing $x$
successes in $n$ independent trials with success probability $p$ (and
understood to be zero unless $0\le x\le n$).  For a more detailed
motivation of (\ref{eq:ydprob}) and (\ref{eq:ycprob}) see \citet[
appendix A]{johnson2005a}

Then the probability of the observed data can be written
\begin{equation}
  \Pr{(\ey{d}{i},\ey{c}{i}|x_i,a_i,\pi_i)} =
  \sum_{\y{d}{i}}{\sum_{\y{c}{i}}{
      \Pr{(\ey{d}{i},\ey{c}{i}|\y{d}{i},\y{c}{i})}
      \Pr{(\y{d}{i}|x_i,a_i,\pi_i)}\Pr{(\y{c}{i}|\pi_i)}
    }}\;\mbox{.}
  \label{eq:SNPi-prob}
\end{equation}

We can simplify at this stage by writing down the distribution
marginal to $a_i$ and $\pi_i$ and moving the respective sum and
integral as far inside the expression as possible, to get
\begin{eqnarray}
  \Pr{(\ey{d}{i},\ey{c}{i}|x_i)} &=&
  \sum_{\y{d}{i}}{\sum_{\y{c}{i}}{\bigg(
      \Pr{(\ey{d}{i},\ey{c}{i}|\y{d}{i},\y{c}{i})}\bigg.}}\nonumber\\
    &&\bigg.\int{\sum_{a_i}{\big(
      \Pr{(\y{d}{i}|x_i,a_i,\pi_i)}\Pr{(a_i|\pi_i)}\big)}\Pr{(\y{c}{i}|\pi_i)}\Pr{(\pi_i)}\dif\pi_i}\bigg)\;\mbox{.}
\end{eqnarray}
The innermost sum over $a_i$ can be rewritten
\begin{eqnarray}
  \label{eq:emprob}
  \Pr{(\y{d}{i}|x_i,\pi_i)} &=&
  \sum_{a_i}{\Pr{(\y{d}{i}|x_i,a_i,\pi_i)}\Pr{(a_i|\pi_i)}} \\
  &=& \sum_{a_i}{\bigg(
    \bindens{(\ya{d}{i}{a_i}-x_i+1,\nd-x_i+1,\pi_i(a_i))}\bigg.}\nonumber\\
  &&\qquad\times
  \bigg.\binom{\nd-x_i}{\ya{d}{i}{a_i}-x_i}\bigg/\binom{\nd-x_i+1}{\ya{d}{i}{a_i}-x_i+1}\bigg)\\
  &=& \sum_{a_i}{\bigg(
    \bindens{(\ya{d}{i}{a_i}-x_i+1,\nd-x_i+1,\pi_i(a_i))}\bigg.}\nonumber\\
  &&\qquad\times
  \bigg.\frac{\ya{d}{i}{a_i}-x_i+1}{\nd-x_i+1}\bigg)\;\mbox{.}
\end{eqnarray}
This allows the integral over $\pi_i$ to be computed
analytically when (as assumed above) the prior $\Pr{(\pi_i(a_i))}$ is a
beta distribution with parameter $(P_{i,a_i},P_{i,1-a_i})$\begin{eqnarray}
  \label{eq:ygivenxraw}
  \Pr{(\y{d}{i},\y{c}{i}|x_i)} &=& \int{
    \Pr{(\y{d}{i}|x_i,\pi_i)}\Pr{(\y{c}{i}|\pi_i)}\Pr{(\pi_i)}\dif \pi_i}\\
  \label{eq:ygivenxbinombeta}
&=&
  \sum_{a_i}{\bigg(\frac{(\nd-x_i+1)!\;\Gamma{(\nc+P_{i,0}+P_{i,1})}}
  {\Gamma{(\nd-x_i+1+\nc+P_{i,0}+P_{i,1})}}\bigg.}\nonumber\\
  &&\qquad\times \frac{\Gamma{(\ya{d}{i}{a_i}-x_i+1+\ya{c}{i}{a_i}+P_{i,a_i})}}
  {(\ya{d}{i}{a_i}-x_i+1)!\;\Gamma{(\ya{c}{i}{a_i}+P_{i,a_i})}}\nonumber\\
  &&\qquad\times \frac{\Gamma{(\ya{d}{i}{1-a_i}+\ya{c}{i}{1-a_i}+P_{i,1-a_i})}}
  {\ya{d}{i}{1-a_i}!\;\Gamma{(\ya{c}{i}{1-a_i}+P_{i,1-a_i})}}\nonumber\\
  &&\qquad\times
  \bigg.\frac{\ya{d}{i}{a_i}-x_i+1}{\nd-x_i+1}\bigg)
  \label{eq:probSNPfull}
\end{eqnarray}
Thus the probability of the observed data reduces to
\begin{equation}
  \Pr{(\ey{d}{i},\ey{c}{i}|x_i)} =
  \sum_{\y{d}{i}}{\sum_{\y{c}{i}}{
      \Pr{(\ey{d}{i},\ey{c}{i}|\y{d}{i},\y{c}{i})}
      \Pr{(\y{d}{i},\y{c}{i}|x_i)}
    }}
  \label{eq:probSNPsumsum}
\end{equation}
where the first term in the summand is the error
model and the second term is given by
(\ref{eq:probSNPfull}).  

Because $\ey{d}{i}$ and $\ey{c}{i}$ are fixed and $a_i$ and $\pi_i$
have been integrated out, values of (\ref{eq:probSNPsumsum}) for each
$x_i$ can be precomputed and stored in a small lookup table of size
$(n_d+1)$.  It is therefore feasible to use a complicated and
hopefully realistic error distribution, as discussed above.  For many
error models, computation can be speeded up without loss of accuracy
by judicious reduction of the range of the summation in
(\ref{eq:probSNPsumsum}).

\subsection{Hidden Markov Model for $x_\mu$ and the $x_i$}
\label{sec:hidden-markov-model}
Throughout this section, all probabilities are conditional on given
values of $\mu$, $\tau$ and $\rho$ but to make a clearer exposition this is
suppressed in the notation.

Assume for clarity of exposition that $\mu$ is a position within the
battery of marker SNPs.  Let $\ceil{\mu}=\min{\{i:m_i\ge \mu\}}$ and
$\floor{\mu}=\max{\{i:m_i< \mu\}}$ denote the indices of the SNPs to
the right and left of the QTL.  The algorithm works with
obvious modifications when $\mu$ is a position outside the battery of
marker SNPs.

Under the modelling assumptions, as we move away from the position of
the disease locus, the number of chromosomes containing i.b.d.\
ancestral chromosome, $x_i$, is Markovian.  The \textit{transition
  probabilities} of a (nonstationary and inhomogenous) hidden Markov
model \cite[HMM; see e.g.][ ch.3]{durbinbook}, for marker SNPs to the right
of $\mu$ (i.e.\ $\ceil{\mu}\le i$), are
\begin{equation}
  \label{eq:jumpprob}
  \Pr{(x_{i+1}|x_{i})}= 
  \bindens{(x_{i+1},\;x_{i},\;\exp{(-\tau\times(m_{i+1}-m_{i}))})}
\end{equation}
Similar equations hold for for marker SNPs to the left of $\mu$, and
for $\Pr{(x_i|x_\mu)}$.  The \textit{emmission probabilities} of the
HMM are given by (\ref{eq:probSNPsumsum}).  There is an important
difference between this HMM and the HMMs of \citet{mcpeek1999},
\citet{morris2000}, \citet{zhangzhao2000,zhangzhao2002} and \citet{liu2001}.  Those
authors modelled the observed haplotypes or genotypes as a set of
conditionally independent HMMs, conditional on (in the present
notation) $(a_1,\ldots,a_L)$, $(\pi_1,\ldots,\pi_L)$ as well as $\mu$,
$\tau$ and $\rho$.  Thus they had to use expensive numerical
algorithms (optimisation or MCMC) on the high dimensional space of
variables on which the HMMs were conditioned.  Here I am able to model
all the data as a \emph{single} HMM conditional on only $\mu$, $\tau$
and $\rho$.  I am thus able to integrate out the high dimensional
$(a_1,\ldots,a_L)$ and $(\pi_1,\ldots,\pi_L)$ using an efficient
numerical algorithm and am left with only a low dimensional space (the
space for $(\mu,\tau,\rho)$) on which I will need to use an expensive
numerical algorithm.

We can use the backwards propagation algorithm for HMMs to sum over
the hidden states $(x_1,\ldots,x_L)$ \citetext{see e.g.\
  \citealt{durbinbook} ch.3, \citealt{liubook2001} pp.28--31}.
Readers familiar with HMMs may wish to skip the rest of this section.

Define the \textit{backwards variables} for SNPs at positions to the
right of the QTL
\begin{equation}
  \label{eq:bvdefright}
  b(x_i) := \Pr{(\ey{d}{i+1},\ey{c}{i+1},\ldots,\ey{d}{L},\ey{c}{L}|x_i)}
\end{equation}
and for SNPs at positions to the left of the QTL
\begin{equation}
\label{eq:bvdefleft}
  b'(x_i) := \Pr{(\ey{d}{1},\ey{c}{1},\ldots,\ey{d}{i-1},\ey{c}{i-1}|x_i)}
\end{equation}
Here backwards is relative to the direction in which the hidden
process is Markovian.  Equation~(\ref{eq:bvdefleft}) should not be
confused with a forwards variable, which are not used in this
computation.  Note also that the backwards variables are functions of
which SNP ($i$) and the value of that $x_i$ at that SNP, but that I have
adopted a more streamlined notation that I hope is unambiguous.

The backwards variables have the obvious interpretation when the
arguments run out of
range, that
\begin{eqnarray}
  b(x_L) &:=& \Pr{(\mbox{nothing}|x_L)} = 1 \\
  \label{eq:hmm-initialise-r}
  b'(x_1) &:=& \Pr{(\mbox{nothing}|x_1)} = 1
  \label{eq:hmm-initialise-l}
\end{eqnarray}

Using (\ref{eq:hmm-initialise-r}) to \textit{initialise} the backwards variables
at $i=L$, we then proceed to \textit{propagate} leftwards along the
chromosomes for each
$i=L-1,\ldots,\ceil{\mu}$ in turn, compute the backwards variables for every $(x_i)$ using
\begin{equation}
  b(x_i) =\sum_{x_{i+1}}{\Pr{(x_{i+1}|x_i)}
    \Pr{(\ey{d}{i+1},\ey{c}{i+1}|x_{i+1})} b(x_{i+1})}
\end{equation}
and then \textit{terminate} the algorithm by computing
\begin{equation}
  \label{eq:modelprobgivenmu}
  \Pr{(\ey{d}{\ceil{\mu}},\ey{c}{\ceil{\mu}},\ldots,\ey{d}{L},\ey{c}{L}|x_\mu)} 
  = \sum_{x_{\ceil{\mu}}}{\Pr{(x_{\ceil{\mu}}|x_\mu)}
    \Pr{(\ey{d}{\ceil{\mu}},\ey{c}{\ceil{\mu}}|x_{\ceil{\mu}})} b{(x_{\ceil{\mu}})}}
\end{equation}
for every $x_\mu$.

Likewise, using (\ref{eq:hmm-initialise-l}) to initialise the $b'$
backwards variables at $i=1$ we can propagate rightwards along the
chromosomes (backwards in the sense that time or space is usually
considered in HMMs) for $i=2,\ldots,\floor{\mu}$ and terminating in a
symmetric manner to compute
$\Pr{(\ey{d}{1},\ey{c}{1},\ldots,\ey{d}{\floor{\mu}},\ey{c}{\floor{\mu}}|x_\mu)}$.

We then obtain the probability of all the data by computing
\begin{eqnarray}
  \label{eq:modelprobgivenrho}
  \Pr{(\hat{\vec{y}})} &=&
  \Pr{(\ey{d}{1},\ey{c}{1},\ldots,\ey{d}{L},\ey{c}{L})} \nonumber \\
  &=& \sum_{x_\mu}{\bigg(\bindens{(x_\mu,n_d,\rho)}
    \Pr{(\ey{d}{1},\ey{c}{1},\ldots,\ey{d}{\floor{\mu}},\ey{c}{\floor{\mu}}|x_\mu)}\bigg.}\nonumber\\
  &&\qquad \bigg. \Pr{(\ey{d}{\ceil{\mu}},\ey{c}{\ceil{\mu}},\ldots,\ey{d}{L},\ey{c}{L}|x_\mu)}\bigg) 
\end{eqnarray}

Restoring the conditioning that has been implicit throughout this
section, (\ref{eq:modelprobgivenrho}) is
$\Pr{(\hat{\vec{y}}|\mu,\tau,\rho)}$, the probability of all the data
conditional on $(\mu,\tau,\rho)$ and marginal to $(a_1,\ldots,a_L)$
and $(\pi_1,\ldots,\pi_L)$,.

\subsection{Marginal Posteriors for $\mu$, $\tau$ and $\rho$ and Model
Probabilities}
\label{sec:final-posterior}
The posterior for $\mu$, $\tau$ and $\rho$, up to a normalising
constant, is obtained simply by multiplying
(\ref{eq:modelprobgivenrho}) by the relevant priors.
\begin{equation}
  \Pr{(\hat{\vec{y}},\mu,\tau,\rho)} = \Pr{(\hat{\vec{y}}|\mu,\tau,\rho)}
  \,\Pr{(\mu)}\,\Pr{(\tau)}\,\betadens{(\rho,R_1,R_0)}
\end{equation}
so
\begin{equation}
  \Pr{(\mu,\tau,\rho|\hat{\vec{y}})} = \Pr{(\hat{\vec{y}}|\mu,\tau,\rho)}
  \,\Pr{(\mu)}\,\Pr{(\tau)}\,\betadens{(\rho,R_1,R_0)}\,\frac{1}{\Pr{(\hat{\vec{y}})}}\;\mbox{.}
\end{equation}
Summarising this posterior is not entirely trivial for large data
sets, because computing the posterior at any given point
$(\mu,\tau,\rho)$ using the propagation algorithm of
section~\ref{sec:hidden-markov-model} takes on the order of
$L\,n_d^2$ operations (and all addition has to be done in log-space,
see e.g.\ \citet[ p.77--78]{durbinbook}), so we cannot rely on being able
to make an arbitrarily large number of such computations.

I use Cartesian product quadrature \citep[CPQ; see e.g.][
\textbf{9.43}--\textbf{9.44}]{ohaganforster} to compute marginal
posteriors for each of $\mu$, $\tau$ or $\rho$, and the normalising
constant or marginal likelihood $\Pr{(\hat{\vec{y}})}$ assuming there
is a disease QTL.  CPQ makes the approximation
\begin{eqnarray}
  \Pr{(\hat{\vec{y}})}&=&\int{\int{\int{
        \Pr{(\hat{\vec{y}},\mu,\tau,\rho)}\,\dif\mu}\,\dif\tau}\,\dif\rho}
  \nonumber\\
  &\simeq&\sum_j{\sum_k{\sum_\ell{ w_j^{(\mu)} w_k^{(\tau)} w_\ell^{(\rho)} \Pr{(\hat{\vec{y}},\mu_j,\tau_k,\rho_\ell)}}}}
  \label{eq:cpq-howto}
\end{eqnarray}
where e.g.\ $j$ indexes a set of \emph{design points}
$\{\mu_1,\mu_2,\ldots\}$, and $w_j^{(\mu)}$ is a \emph{weight}
associated with the $j$-th design point.  (A quantity proportional to)
the marginal posterior for any variable $\mu$, $\tau$ or $\rho$ is obtained by omitting the
respective sum from (\ref{eq:cpq-howto}).  When computed using priors
describing our beliefs about these variables given that there is a
QTL in the region of interest, the quantity in (\ref{eq:cpq-howto})
will be called $\Pr{(\hat{\vec{y}}|\mbox{QTL})}$.

The choice of design points and weights depends on the region of
interest and the prior, and the quadrature rule to be used.  For
example, all calculations in section~\ref{sec:results} the region of
interest is $\mu\in(0,1)$, measured in Mb or cM.  With a uniform prior
for $\mu$, exponential prior for $T$ with $T=1/1000$, and flat beta
prior for $\rho$ with $R_1=R_0=1$, I used 100 design points for each
variable as follows:
\begin{eqnarray}
  \mu_j = \left(j+1/2\right)/100\mbox{,}&
  w_j^{(\mu)} = 0.01\mbox{,}& j=0,1,\ldots,99\nonumber\\ 
  \tau_k = \exp{(k/11)}\mbox{,}&
  w_k^{(\tau)} = \exp{(k/11)}/11\mbox{,}& k=0,1,\ldots,99\nonumber\\
  \rho_\ell = \left(\ell+1/2\right)/100\mbox{,}&
  w_\ell^{(\rho)} = 0.01\mbox{,}& \ell=0,1,\ldots,99
  \label{eq:cpq-default-design}
\end{eqnarray}
Here the quadrature rule is very simple and corresponds approximately
to the trapezoid rule or two point Newton--Cotes rule.

A simple model with no QTL is to assume there are no blocks of i.b.d.\
haplotype, $x_i=0$ for all $i$.  This corresponds to a degenerate
prior at $\rho=0$ (or the limits $\mu\to\pm\infty$ or $\tau\to\infty$).
The probability of the data under this model, $\Pr{(\hat{\vec{y}}|\mbox{no
    QTL})}$, is easily computed directly from
(\ref{eq:probSNPsumsum}).  The Bayes factor (BF) in favour of the
model with a QTL is then
\begin{equation}
  \label{eq:BF-def}
  \mbox{BF} = \frac{\Pr{(\mbox{QTL}|\hat{\vec{y}})}}{\Pr{(\mbox{no
        QTL}|\hat{\vec{y}})}}\bigg/
\frac{\Pr{(\mbox{QTL})}}{\Pr{(\mbox{no
        QTL})}}
=\frac{\Pr{(\hat{\vec{y}}|\mbox{QTL})}}{\Pr{(\hat{\vec{y}}|\mbox{no QTL})}}\;\mbox{.}
\end{equation}
In addition to its Bayesian interpretation, the
BF (or any isotonic transformation thereof) has good
properties, from a classical frequentist perspective, as a test
statistic to test the null model with no QTL against the alternative
with a QTL \citetext{\citealt{ohaganforster} ch.5,
  \citealt{patterson2004}}.  Tests based on the BF are admissible,
which means that no other test has greater power for all
$(\mu,\tau,\rho)$.  There may be other tests that have greater power
for some $(\mu,\tau,\rho)$, and are therefore also admissible, but it
is ``unusual and strange'' to find an admissible test that is not
based on the BF computed using \emph{some} prior.  Furthermore, (up to
isomorphism) the BF uniquely maximises average power, when the
averaging is done with respect to the prior for $(\mu,\tau,\rho)$ used
to compute the BF.  Of course, this theory only applies when the model
is correct, but we might hope that the most powerful test for an
approximate model would be approximately most powerful for a more
realistic model.  The BF is a sensible way to combine information
across many markers to produce a single test, and thus avoids (or
overcomes) the problematic need to correct for multiple testing
\citep{patterson2004}.

I have also implemented a Markov chain Monte Carlo
\citep[MCMC; see e.g.][]{gilks1996,liubook2001} sampler, which uses
the Metropolis--Hastings algorithm to sample from
the posterior density for $(\mu,\tau,\rho)$.  
A proposal distribution that seems to work reasonably well
is to update a single variable, choosen at random with equal
probability, using the following:
\begin{eqnarray}
  \label{eq:proposal-z}
  \mu' &\sim& \ndist{\left(\mu,(0.1(m_L-m_1))^2\right)}\nonumber\\
  \label{eq:proposal-t}
  t' &\sim& \gamdist{\left(10,t/10\right)}\nonumber\\
  \label{eq:proposal-r}
  r' &\sim& \betadist{\left(5r,5(1-r)\right)}
\end{eqnarray}
Kernel based methods \citep[see e.g.][]{silverman1986} can then be
used to estimate marginal posteriors for each of the variables, and
these seemed similar to the marginal posteriors computed using CPQ in
the cases I examined.  However, standard numerical methods to estimate
the model probability $\Pr{(\hat{\vec{y}})}$ from the MCMC output
\cite[see e.g.][ \textbf{10.46}]{ohaganforster} converged very slowly
and did not seem reliable.

In addition to the ease and reliability with which the normalising
constant or BF can be computed, CPQ offers substantial advantages over
alternatives such as MCMC in low dimensional situations such as this
one.  There are no concerns about burnin or mixing.  CPQ makes
efficient use of the evaluation of the posterior density at each
design point. We can investigate sensitivity to prior specification
afterwards without redoing much of the computation.

In this particular situation CPQ offers an additional advantage, that
by traversing the design points in a particular order many of the
backards variables computed in the propagation algorithm of
section~\ref{sec:hidden-markov-model} it be stored and reused, and
thus a CPQ algorithm with $n$ design points runs much faster than a
MCMC algorithm with $n$ samples.  The details are as follows: The
posterior can be computed on all points on a three dimensional lattice
most efficiently by traversing the lattice with different values of
$\tau$ in the outermost loop, different values of $\mu$ in the middle
loop, and different values of $\rho$ in the innermost loop.  This is
because each time $\rho$ changes but $\mu$ and $\tau$ do not then only
(\ref{eq:modelprobgivenrho}) has to be recomputed.  Each time $\mu$
changes but $\tau$ does not then (\ref{eq:modelprobgivenmu}) always
has to be recomputed, and if the old and new values of $\mu$ are
separated by one or more typed SNPs then one or more columns of
backwards variables also have to be recomputed.  If $\mu$ always
increases then the $b$ are all computed first and then as the lattice
is traversed extra columns of $b'$ are computed and columns of $b$ are
simply discarded.  Changing $\tau$ means that the transition
probabilities~(\ref{eq:jumpprob}) change and everything has to be
recomputed.  The run time of the whole algorithm (propagation and CPQ)
scales quadratically in $n_d$ and linearly in $L+d_\mu$ where $d_\mu$
is the number of design points used for $\mu$.  If a small map region
is found to be interesting additional design points can be added later
at moderate cost.

The disadvantages of CPQ are that we learn little about the posterior
until calculations have been completed for all the design points, we
may belatedly discover that our choice of design points was not a good
one, that MCMC \emph{may} be more sensitive to very narrow spikes
containing substantial probability mass (these will be missed if they
fall inbetween the design points) and that CPQ will not scale well to
higher dimensional spaces which we might need to study if we
elaborated the model.

\section{Coalescent Simulation Model and Error Model}
\label{sec:simul-model}
In this section I describe a more realistic model that was used to
simulate datasets on which to test the method described above in
sections~\ref{sec:appr-model-prior}--\ref{sec:analysis}.  I also
describe a specific model for errors in allele frequency estimation.
Each simulated dataset was generated as follows.

First I used the \texttt{mksamples} program of \citet{hudson2002} to
simulate a sample of 20,000 1Mb long regions, assuming the standard
neutral coalescent model with population recombination rate
$4N_{\mathrm{e}}c=400$ ($N_{\mathrm{e}}=10,000$ assuming
$1\mbox{cM}/\mbox{Mb}$), and assuming the infinite sites mutation
model with population mutation rate $4N_{\mathrm{e}}\mu=10$.  (This is
an unrealistically low mutation rate, the idea is to simulate some of
the SNPs in the region rather than all of them.)  Chromosomes were
paired at random to generate a sample of 10,000 individuals.

One SNP with a minor allele frequency between 10\% and 20\% was chosen
as the disease QTL.  The disease status of each individual was
simulated assuming multiplicative risks, so that
$\gamma_{01}/\gamma_{00}=\gamma_{11}/\gamma_{01}=g$.  Here $\gamma_G$
is the penetrance (probability of having the disease) given genotype
$G$ at the disease QTL, with 1 the minor allele.  The parameter $g$ is
called the allelic or genotype relative risk 
\cite[see e.g.][]{rischmerikangas1996}.  Simulations for this paper used
values of $g=4$ and $g=1$.  The penetrance of the wild type
homozygote, $\gamma_{00}$, was set so that the marginal probability of
having the disease was 0.02.  (Thus the number of case chromosomes
$\nd$ was random with expectation $0.02\times10,000\times2=400$.)
Data from all $\nd/2$ case individuals, and an equal number $\nc/2$ of
randomly chosed control individuals, were used to make up the two
pools.

Excluding the disease QTL, all simulated SNPs with a minor allele
frequency greater than 0.05 in the $\nc/2$ individuals in the control
pool were analysed, so the number of SNPs $L$ was also random.  The
estimated allele frequencies at each SNP and for each pool were either
assumed to be known exactly, or assuming that allele frequencies were
estimated using the lag between kinetic PCR growth curves
\citep{germer2000}, using
\begin{equation}
  \label{eq:yhatfromlag}
  \hat{y} = \frac{1}{1+2^{\Delta\widehat{C_t}}}\times n\;\mbox{.}
\end{equation}
Here $\hat{y}$ is shorthand for the estimated count of the 1 allele,
$n$ is the number of chromosomes in the pool,
$\Delta\widehat{C_t}=\widehat{C_t}(1)-\widehat{C_t}(0)$, and
$\widehat{C_t}(a)$ is the number of PCR cycles before the amount of
PCR product for allele $a$ exceeds some threshold level
\citep{germer2000}.  The model for $\Pr{(\hat{y}|y,e)}$ is then as
follows: Define the true lag $\Delta C_t=\log_2((n-y)/y)$, which is
the lag that would give the correct frequency when
(\ref{eq:yhatfromlag}) was used.  I assume that the observed lag
$\Delta\widehat{C_t}$ averaged across $r$ experiments is normally
distributed with mean $\Delta C_t$ and variance $\sigma^2/r$, where
$\sigma^2$ is the variance in lags across replicate experiments.

Using the Jacobian
\begin{equation}
  \label{eq:yhatzhatjacobian}
  \frac{\dif \hat{y}}{\dif (\Delta\widehat{C_t})} = \ln{(2)}\hat{y}(n-\hat{y})\frac{1}{n}
\end{equation}
we can write down the error model in
the form required in (\ref{eq:SNPi-prob}) for the analysis,
\begin{equation}
  \label{eq:errorgaussdeltact}
  \Pr{(\hat{y}|y,n,\sigma^2,r)} = \frac{n}{\ln{(2)}\hat{y}(n-\hat{y})}
  \frac{1}{\sqrt{2\pi\sigma^2/r}}
  \exp{\left(-\frac{\left(\ln\left(\frac{(n-y)\hat{y}}{y(n-
\hat{y})}\right)\right)^2}{2\ln{(2)}^2\sigma^2/r}\right)} \;\mbox{.}
\end{equation}

\section{Testing the Method} 
\label{sec:results}
All measurements of the performance of the Bayesian method described
in sections \ref{sec:appr-model-prior}--\ref{sec:analysis} are based
on analyses of datasets simulated under a more realistic model, as described in
section~\ref{sec:simul-model}.  Results are reported for two situations, either where the allele
frequencies in each pool are known exactly, or where there are errors in allele
frequency estimation using (\ref{eq:errorgaussdeltact}) and assuming
that $n$, $r=2$ replicates and $\sigma=0.2\mbox{ PCR cycles}$ are all known.
This magnitude of error is comparable to those reported by
\citet{germer2000} and \citet{shiffman2004}.
Using (\ref{eq:yhatzhatjacobian}) we can say that these parameter values
correspond to a ``typical'' error in allele frequency estimate
$\hat{y}/n$ of about $y/n (1-y/n)\ln{(2)}\sigma/\sqrt{2}\simeq y/n
(1-y/n)0.098$ or, for intermediate allele frequencies, about 2.5\%.

For each situation, allele frequencies known either exactly or
estimated with errors, I analysed 500 datasets simulated
assuming there was a QTL with a
genotype relative risk $g=4$, and 500 datasets simulated assuming a null model
with no QTL ($g=1$, so the penetrances
$\gamma_{00}=\gamma_{01}=\gamma_{11}=0.02$ are all equal).  In these
simulations, the median number of case or control individuals,
$\nd/2=\nc/2$, was 200 (interquartile range 191--209, range 154--248).
The median number of SNPs, $L$, was 28 (interquartile range 24--32,
range 12--51).  These simulations assumed relative risks that are 
higher, and correspondingly sample sizes that are smaller than may be
realistic for many studies of QTLs influencing complex genetic
diseases.  This reflects the need to analyse a reasonably large number
of simulated datasets with the computing resources currently available
to me.  The mean time to run an analysis on a simulated dataset, using
CPQ with the design (\ref{eq:cpq-default-design}) which requires
evaluating the posterior at $10^6$ points on a $100\times100\times100$
lattice, was about 36 minutes on a 2.4GHz Intel\textregistered{}
Xeon\texttrademark{} processor (totalling about 50 processor days for
all the simulated datasets).

The inferences from the Bayesian method described here are compared
against simple but widely used classical single point analyses.  When
allele frequencies in each pool are known exactly, a chi squared test
can be used on the counts of the two alleles in the two pools
\citep[see e.g.\ ][]{clayton2001}, at each
marker SNP separately.  \citet{visscher2003} consider how to perform
an equivalent test when the errors in allele frequency estimates are
Gaussian.  The relatively small Gaussian errors in $\Delta\widehat{C_t}$
simulated here will produce errors in allele frequency estimates that
are approximately Gaussian (to the extent that (\ref{eq:yhatfromlag})
is linear, and in fact are underdispersed relative to a Gaussian in
the direction of extreme allele frequencies).  \citet{visscher2003}
show that a ``shrunk'' test statistic is approximately distributed as
$\chi^2_{(1)}$.  This shrunk statistic is equal to the ordinary
$\chi^2$ statistic computed using a point estimate of the counts,
times a factor $2V_s/(2V_s+V_e)$ where $V_s$ is the estimated sampling
variance of the allele frequency due to sampling a finite number of
cases and controls, under the null hypothesis of equal allele
frequency in cases and controls, and $V_e$ is the variance of the
allele frequency in either pool due to experimental error.  Using
(\ref{eq:yhatzhatjacobian}), for the simulations performed here this
shrinking factor is (approximately, for small $\sigma$)
\begin{equation}
  \frac{2}{2+(\nd+\nc)\hat{p}(1-\hat{p})\ln{(2)}^2\sigma^2/r}
  \;\mbox{.}
  \label{eq:visscher-shrink-factor}
\end{equation}
where $\hat{p}$ is the allele frequency estimated under the null
hypothesis, i.e.\ by pooling the case and control pools.

The most widely considered statistics from a classical single point
analysis are as follows:  Let \pmin{} be the smallest $p$-value of the
$L$ (shrunk) chi squared tests applied to a given dataset, and let
\muminp{} be the map position of the marker with the smallest
$p$-value.

It is worth emphasising that all the tests described in the following
sections concern the classical sense performance of statistics
computed from the Bayesian analysis.  Strictly, the Bayesian sense
performance can only be tested by conducting a Bayesian analysis
assuming a more realistic model (or prior).

\subsection{Power to Detect a QTL}
\label{sec:power-detect-qtl}
\begin{figure}[p]
  \begin{center}
    \includegraphics{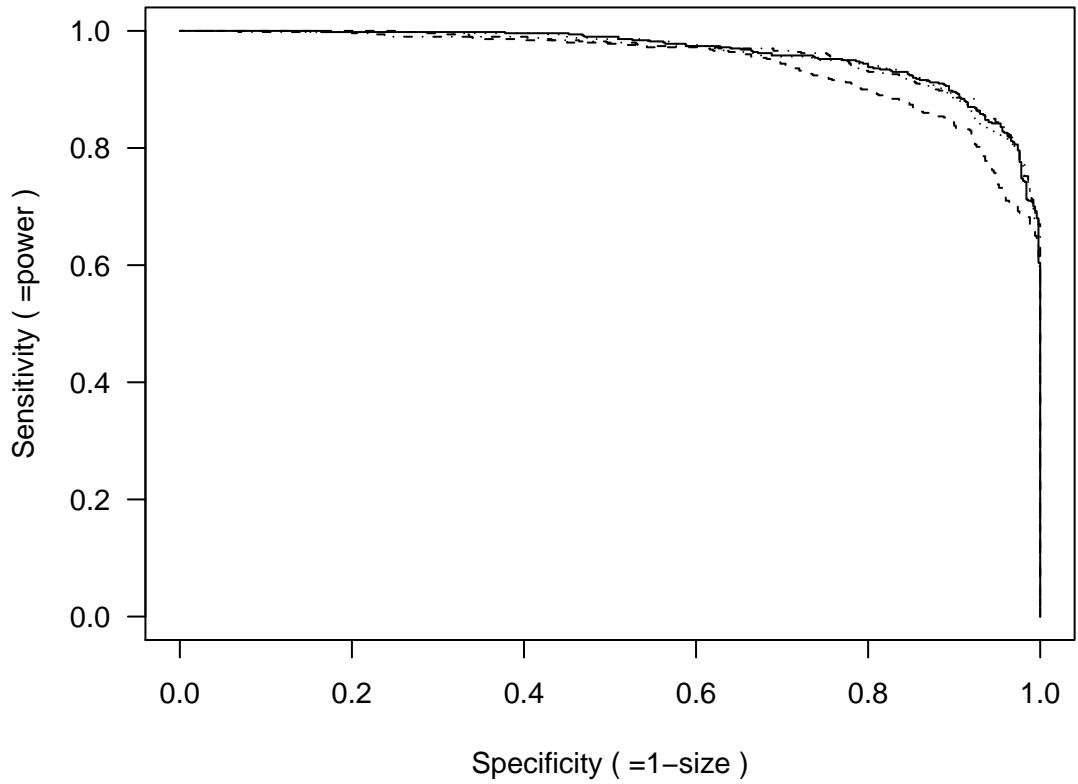}
  \end{center}
  \caption{Sensitivity vs.\ specificity for $2\ln\mbox{BF}$ (solid
    line) and $\pmin\times L$ (dashed line), when allele frequencies
    in each pool are known exactly.  The dotted line shows the
    performance of $2\ln\mbox{BF}$ computed using the priors obtained
    from figure~\ref{fig:new-prior}, and the dot-dashed line shows the
    performance of a nonparametric likelihood ratio test
    statistic \citep[; see text]{johnson2005a}.}
  \label{fig:roc-no-error}
\end{figure}

\begin{figure}[p]
  \begin{center}
    \includegraphics{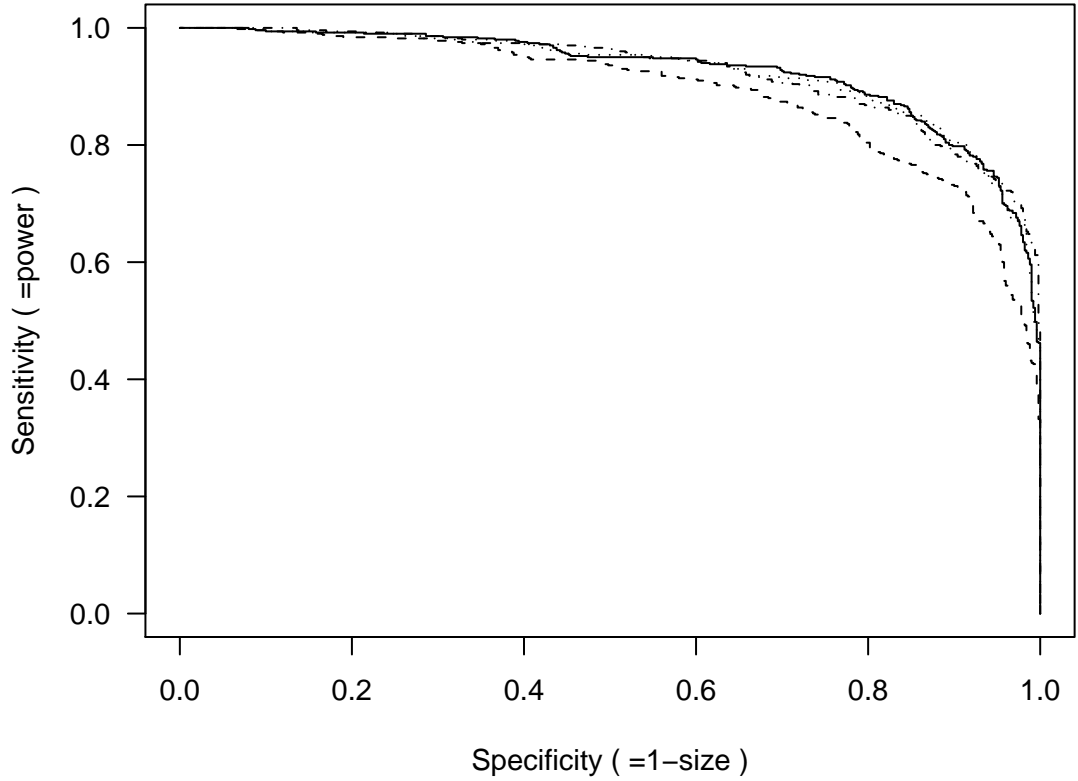}
  \end{center}
  \caption{Sensitivity vs.\ specificity for $2\ln\mbox{BF}$ (solid
    line) and $\pmin\times L$ (dashed line), when there are errors in
    allele frequency estimation with $r=2$ replicates and
    $\sigma=0.2\mbox{ PCR cycles}$.  The dotted line shows the
    performance of $2\ln\mbox{BF}$ computed using the priors obtained
    from figure~\ref{fig:new-prior}, and the dot-dashed line shows the
    performance of a nonparametric likelihood ratio test statistic
    \citep[; see text]{johnson2005a}.}
  \label{fig:roc-with-error}
\end{figure}

In this section I compare the power of tests to detect a QTL.  I
consider two different test statistics, and different methods of
determining critical regions.  The first test statistic is
$2\ln\mbox{BF}$, twice the logarithm of the Bayes factor
(\ref{eq:BF-def}).  The second test statistic is $\pmin\times L$.
Multiplying the smallest single point $p$-value by $L$ achieves a
simple Bonferonni correction for multiple testing that makes the
critical region independent of $L$.  Critical regions were determined
either analytically (by arbitrary or approximate methods), or
empirically (from analyses of datasets simulated under the null
model).  I report the performance of tests with nominal sizes of
$\alpha=0.05$ and $\alpha=0.01$; a more general comparison is made in
figures~\ref{fig:roc-no-error} and \ref{fig:roc-with-error}.  For each
test, the true size was estimated using 500 simulations under the null
model with genotype relative risk $g=1$, i.e.\ where risk is
independent of genotype at the QTL.  The power against an alternative
with $g=4$ was estimated using 500 simulations.  For each test I
report the estimated size and power, along with exact 95\% binomial
confidence intervals for their values.

From a Bayesian perspective, $2\ln\mbox{BF}>0$ indicates evidence in
favour of the model with a QTL over the model with no QTL.  As 
tables~\ref{tab:power-no-error} and \ref{tab:power-error}
show, the test with this critical region has small size and reasonable
power.  However, much more simulation work, for different models and
combinations of parameters, is required to establish the generality of
this result.  Also, at least from a classical perspective it is
desirable to be able to choose a critical region according to the size
(or perhaps power) that is desired.

An arbitrary critical region is
$2\ln\mbox{BF}>2\ln{(\frac{1-\alpha}{\alpha})}$.  I say arbitrary
because this in fact guarantees nothing about the classical sense
error rate, but does bound the Bayesian sense error rate: The
posterior probability that there is no QTL is less than $\alpha$,
$\Pr{(\mbox{no QTL}|\hat{y})}<\alpha$, when the model, the prior
$\Pr{(\mbox{QTL})}=\Pr{(\mbox{no QTL})}$, and the prior for
$(\mu,\tau,\rho)$ are correctly specified. It can be seen from
tables~\ref{tab:power-no-error} and \ref{tab:power-error} that these
arbitrary critical regions give tests with true sizes that are smaller
than $\alpha$.  Such tests are therefore conservative.  Causes may
include the simplified model used to compute the BF, the dependence of
the BF on the prior specification $T$, and the fact that the critical
region bounds the Bayesian sense error rate rather than controls the
classical sense error rate.  The use of these arbitrary critical
regions $2\ln\mbox{BF}>2\ln{(\frac{1-\alpha}{\alpha})}$ entails a loss
of power due to the actual size of the test being smaller than
intended, so a better method for determining a critical region is
desirable.

Assuming goodness of the $\chi^2_{(1)}$ approximation, with the shrinking
factor (\ref{eq:visscher-shrink-factor}), and using simple Bonferonni
correction, suggests an approximate critical region $\pmin\times
L<\alpha$.  These tests have true sizes equal to or slightly smaller
than their nominal sizes, which is expected because the Bonferonni
correction is conservative.  This effect is expected to increase in
severity as the marker density increases, because there will be a
greater number of more positively correlated tests.

Although these respectively arbitrary and approximate methods for
determining critical regions are not terribly accurate, and cannot be
recommended, it is worth noting that there is no clear difference in
power between the two test statistics, for tests with the same nominal
size.  Since the test based on $2\ln\mbox{BF}$ is more conservative,
it might reasonably be preferred.

It is not very meaningful to compare the power of tests with different
sizes.  Therefore I used simulations to estimate exact critical
regions, so that the power of different tests with true size $\alpha$
could be compared.  For these tests, the estimated size is exactly
equal to the nominal size, because the same set of simulations are
used to compute both values.  Although the critical region for
$\alpha=0.01$ is unlikely to be well estimated using only 500
simulations, by a simple symmetry argument this procedure still allows
a fair comparison across the different test statistics.  In every case
the test based on $2\ln\mbox{BF}$ is substantially more powerful than
the test based on $\pmin\times L$.

By combining simulations in which there is not and is a QTL, we can
view test statistics as \emph{classifiers}, and ask how well they
discriminate between the two cases.  The receiver operating
characteristics (ROC) for the two statistics are compared in
figures~\ref{fig:roc-no-error} and \ref{fig:roc-with-error}.  The ROC curves are equivalent to plotting estimated power
($=\mbox{sensitivity}$) against size of test ($=1-\mbox{specificity}$)
for all possible tests (in fact, only all tests with non-disjoint
critical regions).  When viewed in this way, it can be seen that the
BF based statistic is uniformly equal to or superior to the minimum
$p$-value based statistic.  The advantage of the BF is greater when
there are errors in allele frequency estimation.  This may be because,
when the dataset is less informative, it may be more important to have
a model based way to combine information across SNPs.

For comparison, I have also plotted the ROC curves for a test
statistic derived from the nonparametric likelihood approach that I
have described previously \citep{johnson2005a}.  This nonparametric
likelihood ratio (NLR) statistic is defined in the same way as the
BF~(\ref{eq:BF-def}), using the same value for
$\Pr{(\hat{\vec{y}}|\mbox{no QTL})}$, but makes the approximation
\begin{equation}
  \label{eq:proftestdef}
  \Pr{(\hat{\vec{y}}|\mbox{QTL})} \simeq
  \prod_{i=1}^{L}{\Pr{(\ey{d}{i},\ey{c}{i}|x_i^*)}}
\end{equation}
where $(x_1^*,x_2^*,\ldots,x_L^*)$ is the set of hidden states in the
HMM that maximise the probability~(\ref{eq:proftestdef}) under an
order restriction that they are either a weakly increasing sequence, a
weakly decreasing sequence, or a weakly increasing then weakly
decreasing sequence.  This order restriction must be true regardless
of the shape of the genealogy at the QTL.  The NLR statistic is not a
very good approximation to the BF, in particular because it can never
be negative.  As far as I know, there is no theoretical reason to
believe that it should have good properties as a test statistic.
However, as figures~\ref{fig:roc-no-error} and
\ref{fig:roc-with-error} show, tests based on the NLR are superior to
tests based on $\pmin\times L$ and are not clearly distinguishable
from tests based on the BF.  For the simulated datasets studied here,
once the lookup table of emmission
probabilities~(\ref{eq:probSNPsumsum}) has been computed, computing
the NLR is over $10^4$ times faster than computing the BF.
Furthermore, a Viterbi-like algorithm \citep[see][]{durbinbook} for
computing the NLR has time complexity $\mathrm{O}{(L\times\nd)}$,
compared with the CPQ algorithm for computing the BF which has time
complexity $\mathrm{O}{((L+d_\mu)\times\nd^2)}$.

It may concern some readers that the critical regions and sizes and
powers of tests were all estimated while allowing the numbers of
cases, controls and marker SNPs all to vary across simulations.  To
interpret results acquired in this way, a formal classical framework
would require us to view the genotype relative risk $g$ as the single
parameter, and the number of case chromosomes \nd{} and number of SNPs
$L$ as random variables.  It is true that even in such a framework we
would normally wish to perform tests conditional on the values of
ancilliary variables such as \nd{} and $L$ that contain no information
about whether there is a QTL in the region.
However, it is a feature (or weakness) of classical inference that one
is often free to choose whether to condition on any given variable
\citep[but see e.g.][]{jaynes1976,robinson1979}.  The present results
therefore do have a sound classical interpretation.  In any case, the
small number of simulations performed here do not allow the luxury of
estimating critical regions conditional on \nd{} or $L$.  The
simulation procedure as used reflects a likely feature of real
datasets, that SNP density will be higher in regions of the genome
where the genealogy is deeper.  To alter the simulation procedure so
that all simulated datasets had the same value of $L$ would require
the introduction of an \textit{ad hoc} algorithm to select the $L$
markers to be used from a larger number of candidates.

\begin{figure}[p]
  \begin{center}
    \includegraphics{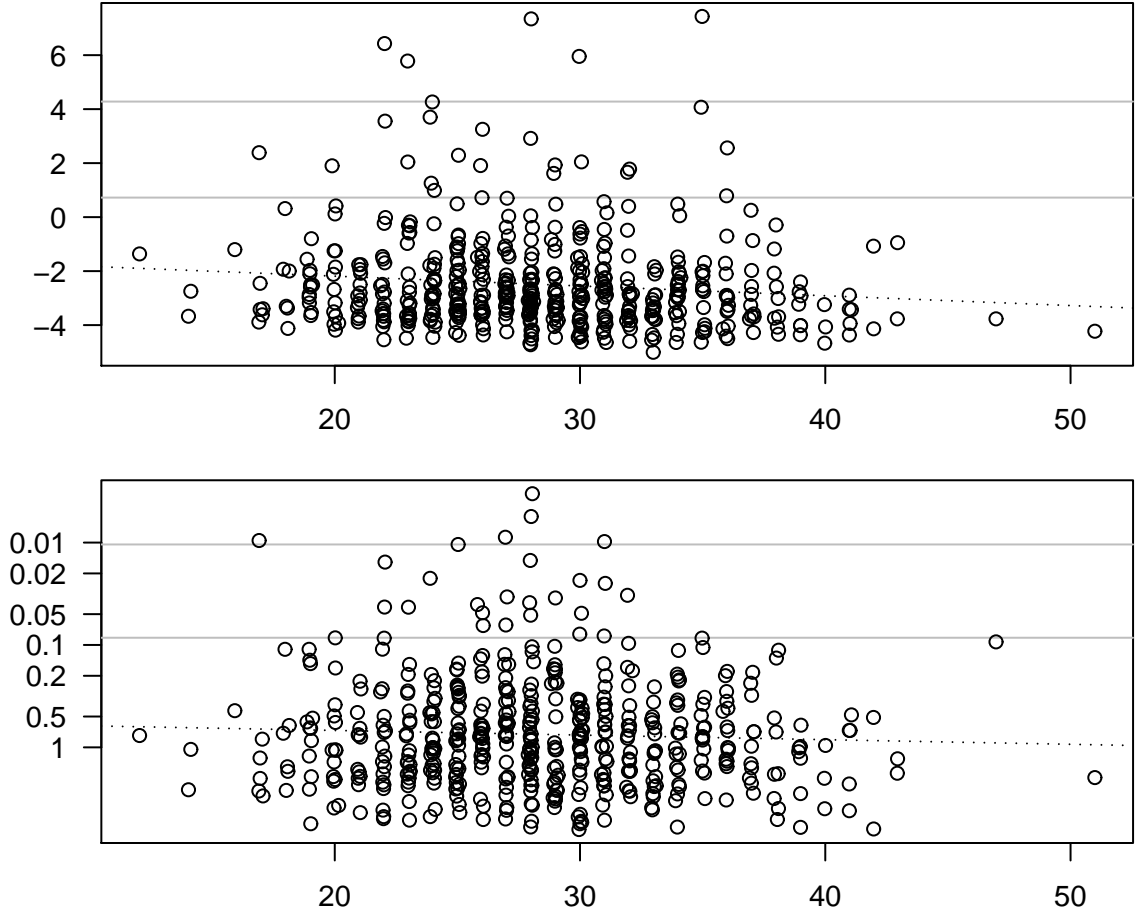}
  \end{center}
  \caption{Sampling distribution of test statistics $2\ln\mbox{BF}$
    (top) and $\pmin\times L$ (bottom, on log scale) under null model
    ($g=1$), as functions of $L$, the number of SNPs in the simulated
    data set.  The 0.95 and 0.99 quantiles are shown as solid lines.
    The least squares linear regression is shown as a dotted line.
    Results shown are for the situation where there are errors in
    allele frequency estimation, but results are similar when allele
    frequencies are known exactly.}
  \label{fig:teststats}
\end{figure}

As shown in figure~\ref{fig:teststats}, the null distributions of both test statistics show a negative
relationship with $L$.  The negative relationship is most pronounced
for the $2\ln\mbox{BF}$ statistic, for the situation where there are
errors in allele frequency estimation.  In this case a linear
regression of $2\ln\mbox{BF}$ on $L$ had a slope significantly
different from zero ($p=0.010$), and if the values of $2\ln\mbox{BF}$
are partitioned into two groups according to the rank of $L$, the
hypothesis that they are drawn from the same distribution can be
rejected using a Kolmogorov--Smirnov test ($p=0.004$).  These tests do
not detect significant dependence ($p>0.05$) for the $2\ln\mbox{BF}$
statistic when the allele frequencies are known exactly, or for the
$\pmin\times L$ statistic.

Although the simulations described here are adequate for demonstrating
the superiority of the BF based test over the \pmin{} based test, we
should be cautious about extrapolating from the current results.  In
particular, it seems that the arbitrary
($2\ln\mbox{BF}>2\ln{(\frac{1-\alpha}{\alpha})}$) or approximate
(Bonferonni) critical regions described above will become more
conservative as SNP number or density increases.  Performing tests
that are not conditioned on SNP number and density will introduce
recognisable subset biases \citep[see e.g.][]{robinson1979}.  In a
real situation, a critical region should be determined using
simulations conditioned on as many ancilliary statistics of the
observed data as possible, although for complex simulation models it
may be a matter of guesswork which statistics are approximately
ancilliary.  An approach that could be most useful in practice is a
variant of the permutation test of \citet{churchill1994}.  This could
be applied if there were matched pairs of pools of cases and controls,
and each pair were typed in separate DNA pooling experiments
\citep{shiffman2004}.  Then the phenotype labels could be permuted
within each pair, giving a set of equiprobable values for any test
statistic under the null hypothesis.  Such an approach could not be
explored here because it would require too much computation.

\subsection{Sensitivity to prior specification}
\label{sec:prior-specification}
It is important to appreciate that the Bayes factor does not depend on
the prior probabilities for the two models (QTL or no QTL), but
\emph{does} depend on the priors for the parameters within the QTL
model.  Misspecification of these priors could adversely influence the
performance of the BF as a test statistic, and it is important to
examine typical levels of robustness to the prior.  To explore this, I
compare the analyses above that used relatively flat generic priors to
analyses that used priors that were in a way optimised for the
simulated datasets under consideration.

\begin{figure}[p]
  \begin{center}
    \includegraphics{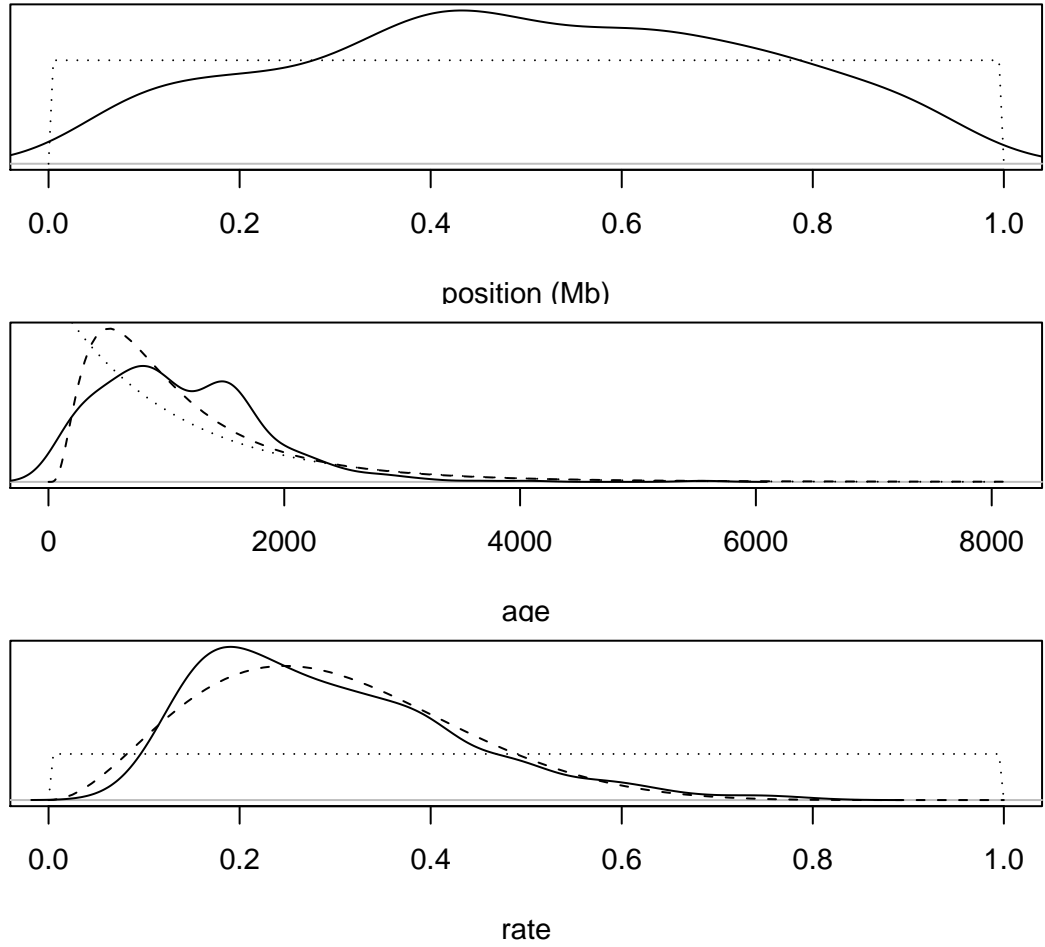}
  \end{center}
  \caption{Original priors (dotted lines) and distribution of
    posterior expectations (solid lines) for the three parameters of
    the approximate model.  This suggests more accurately specified
    priors (dashed lines) as described in the text.}
  \label{fig:new-prior}
\end{figure}

Note that the prior for $\mu$ is correct, but that the approximate
model here uses two other parameters $\tau$ and $\rho$ that do not
have any direct correspondance to parameters of the coalescent model
that the data were simulated under.  It is therefore difficult to say
what the best prior is for analysing the simulated datasets.  Loosely
speaking, we might imagine that for any one simulated dataset, in the
limit of an infinite amount of informative data the posterior for
$\tau$ or $\rho$ would converge to a single value, which we could call
the ``best approximating'' value for that dataset.  However, with less
than an infinite amount of data the posterior mean for either variable
would lie somewhere between the prior mean and the best approximating
value.  Thus the distribution of posterior means across simulations
would be (very loosely speaking) inbetween the degenerate distribution
at the prior mean, and the distribution of best approximating values.
Figure~\ref{fig:new-prior} shows that this is indeed the case for $\mu$, for which the prior mean
is 0.5 and the true correct prior is uniform on $[0,1]$.  The
distributions of posterior means for $\tau$ and $\rho$ shown in
figure~\ref{fig:new-prior} suggests a lognormal prior for $\tau$ (with
$\ln{(\tau)}$ having prior mean 6.8 and prior standard deviation 0.74)
and a beta prior for $\rho$ (with $R_1=3.2$ and $R_0=7.8$, $\rho$
having prior mean 0.29).  Here I am assuming independent priors.  Note
that this exercise in prior specification was totally \textit{ad hoc}.

As can be seen in figures~\ref{fig:roc-no-error} and
\ref{fig:roc-with-error}, the ROC of the test statistic
$2\ln\mbox{BF}$ computed using these priors is hardly 
different better than that using the original priors.  The power for
tests of sizes $\alpha=0.05$ and $\alpha=0.01$ is not significantly
different, based on 500 simulations.  This suggests that the
performance of the BF as a test statistic, for these datasets, is
quite robust to prior specification within the QTL model.

Because most of the computation in QPQ can be reused, computing the BF
for a different prior took on average less than three minutes,
compared with the 36 minutes required to compute the BF for the
original prior.

\subsection{Estimation of QTL Position}
\label{sec:estim-qtl-posit}
\begin{figure}[p]
  \begin{center}
    \includegraphics{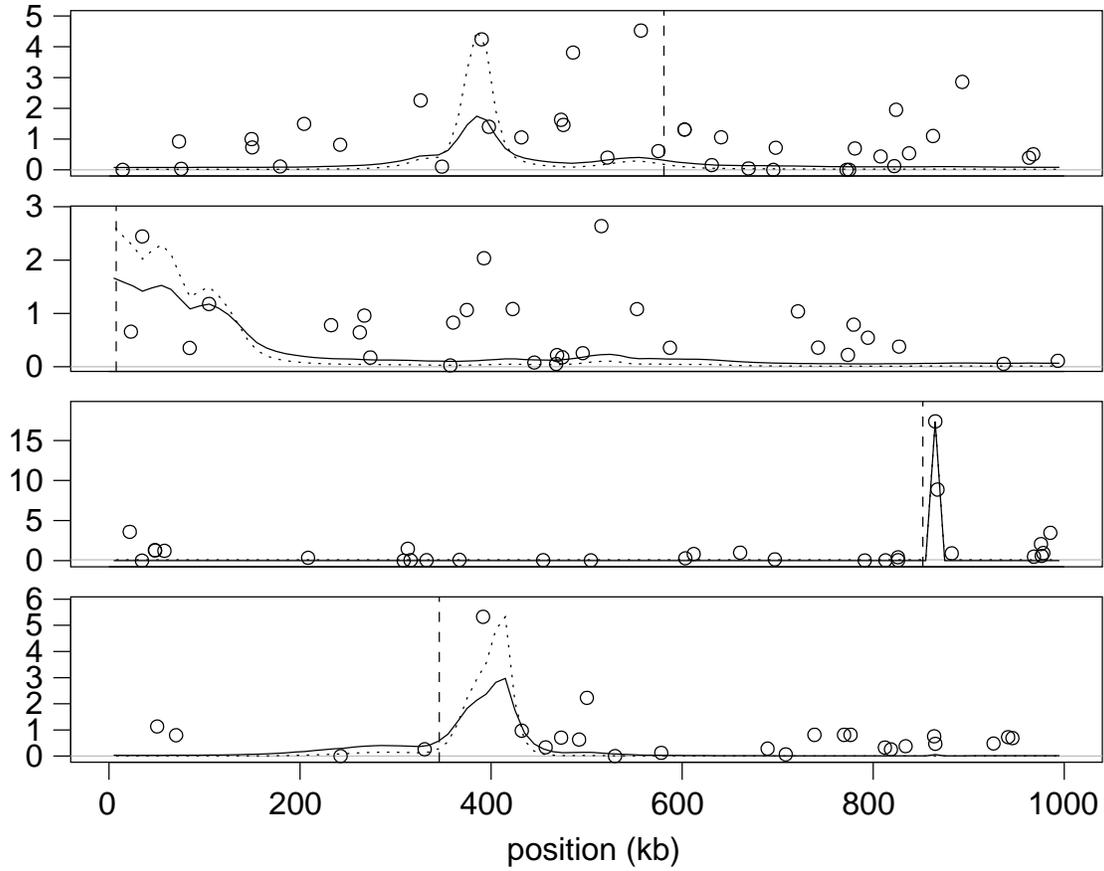}
  \end{center}
  \caption{Example simulated datasets with $g=4$ and where allele
    frequences are known exactly.  Points are $-\log_{10}{p}$ for
    single point $\chi^2$ tests.  Dotted lines are posterior density
    and solid lines are posterior density with McPeek--Strahs
    correction.  Vertical dashed lines show position of disease QTL.}
  \label{fig:example}
\end{figure}

\begin{figure}[p]
  \begin{center}
    \includegraphics{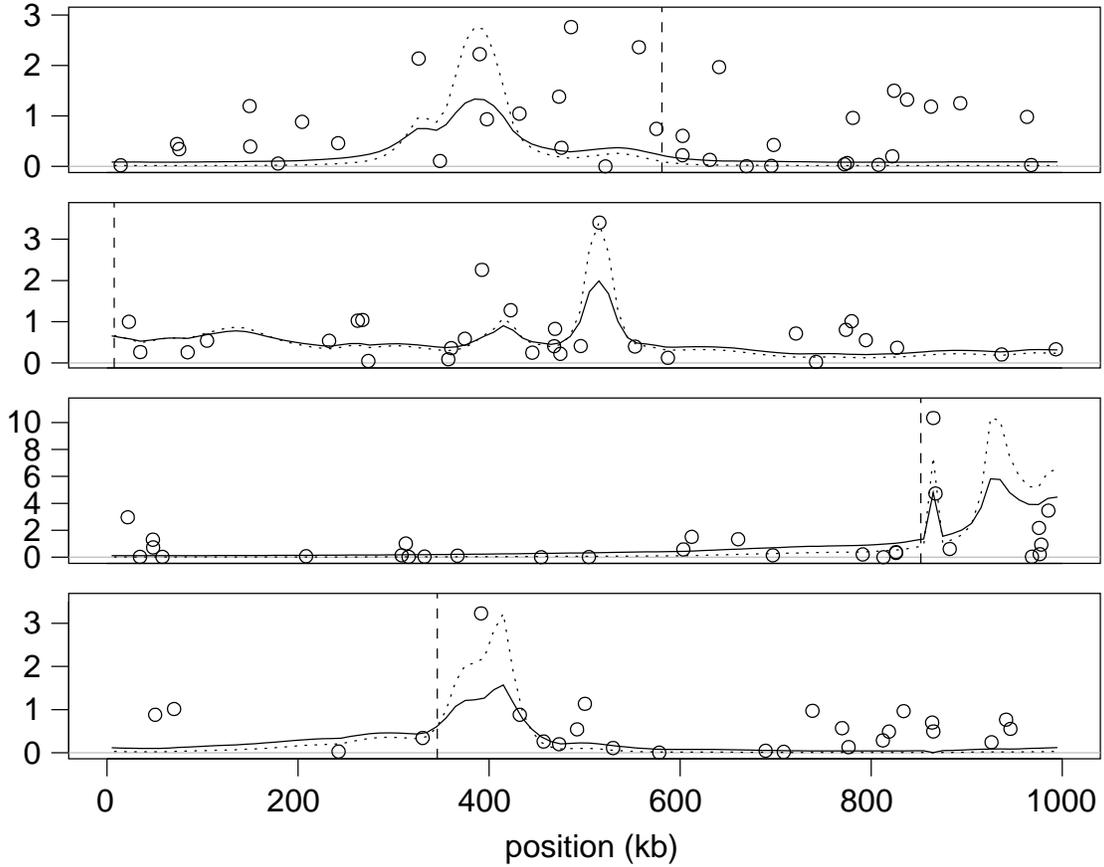}
  \end{center}
  \caption{The same simulated datasets as shown in
    figure~\ref{fig:example}, but with errors in allele frequency
    estimation with $\sigma=0.2$ PCR cycles and $r=2$ experimental
    replicates.  Points are $-\log_{10}{p}$ for single point shrunk
    \citep{visscher2003} $\chi^2$ tests.  Dotted lines are posterior
    density and solid lines are posterior density with McPeek--Strahs
    correction.  Vertical dashed lines show position of disease QTL.}
  \label{fig:nexample}
\end{figure}

Figures~\ref{fig:example} and \ref{fig:nexample} show analyses of four
randomly chosen simulated datasets with QTLs ($g=4$).  These
illustrate the fact that these datasets contain only weak information
about the position of the QTL (or at least that the Bayesian method
described here only extracts weak information).  It nonetheless seems
worthwhile to examine how much information is present.

It has been suggested that the map position of the marker with the
most significant single point test result (i.e.\ the minimum
$p$-value) would be a ``good'' point estimate for the position of the
QTL \citep{kaplanmorris2001a,kaplanmorris2001b}.  However, I point out
that it is asymptotically inadmissible for a model very similar to the
one assumed here.  This argument considers the limit of a QTL of small
effect.  One can imagine models where the position of the marker with
the minimum $p$-value, \muminp, will be tend to become uniformly
distributed on $(0,1)$, independent of the true value of $\mu$, as the
effect of the QTL tends to zero.  The estimator \muminp{} then has
expected loss $\mu(1-\mu)+\frac{1}{2}$ under absolute error loss and
$\frac{1}{3}-\mu(1-\mu)$ under squared error loss.  The estimator
$\hat{\mu}=\frac{1}{2}$ has uniformly lower expected loss,
$|\frac{1}{2}-\mu|$ under absolute error loss and
$(\frac{1}{2}-\mu)^2$ under squared error loss.  This argument does
not technically apply for the model simulated here because SNPs
(including the QTL) tend to be concentrated in regions where the
genealogy is deepest, so even completely ignoring the genotype data,
the position of any SNP is informative about the positions of all
other SNPs including the QTL.  It does however suggest that better
point estimates may be found, and suggests what their asymptotic
behaviour ought to be, at least approximately.

The performance of different methods for estimating the position of
the QTL was assessed using the 500 simulations with $g=4$, for the two
situations with and without errors in allele frequency estimation.
Due to the nature of the simulations performed here, the errors
reported are averaged over the distribution of the true value of
$\mu$.  They are therefore not classical expected losses in the strict
sense, but expected losses averaged with respect to a distribution of
parameter values.  Bayesian point estimators have uniquely best
performance when measured in this way \citep[ ch.5]{ohaganforster};
the theory requires that the model and prior are both correct.  In
particular, under squared error losses the average expected loss is
minimised by the posterior mean, and under absolute error losses the
average expected loss is minimised by the posterior median.  As shown
in table~\ref{tab:point-exp-loss}, point estimators derived from the
posterior calculated using the Bayesian method described here are
superior to \muminp{}, the map location of the marker with the
smallest $p$-value.  When allele frequencies in each pool are known
exactly the Baysian analysis produces a 21\% reduction in root average
mean squared error and a 13\% reduction in average mean absolute
error. When there are errors in allele frequency estimation the
figures are similar, 18\% and 11\% respectively.  The nonparametric
method developed previously by me \citep{johnson2005a} produces point
estimates that are competitive with the Bayesian method under squared
error losses.

\begin{figure}[p]
  \begin{center}
    \includegraphics{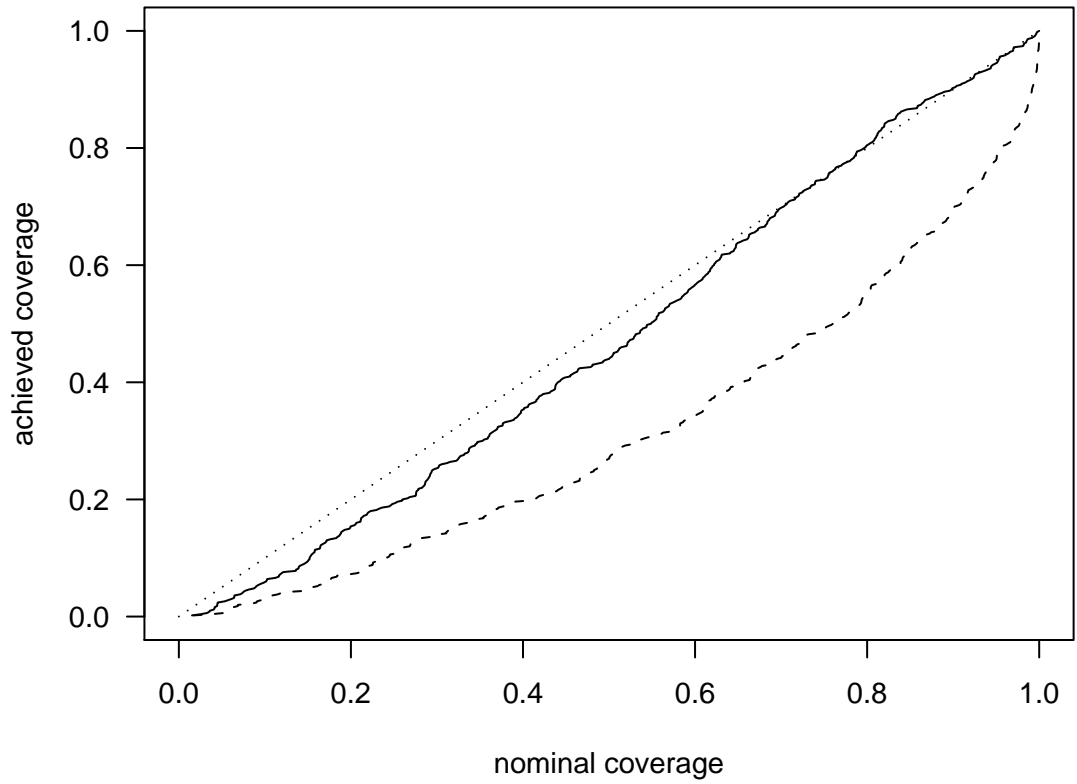}
  \end{center}
  \caption{Nominal and achieved coverage of credibility intervals for
    position of QTL, when allele frequencies are known exactly.
    Credibility intervals were constructed either without (dotted
    line) or with (solid line) the approximate correction factor of
    \citet{mcpeek1999}.}
  \label{fig:coverage}
\end{figure}

\begin{figure}[p]
  \begin{center}
    \includegraphics{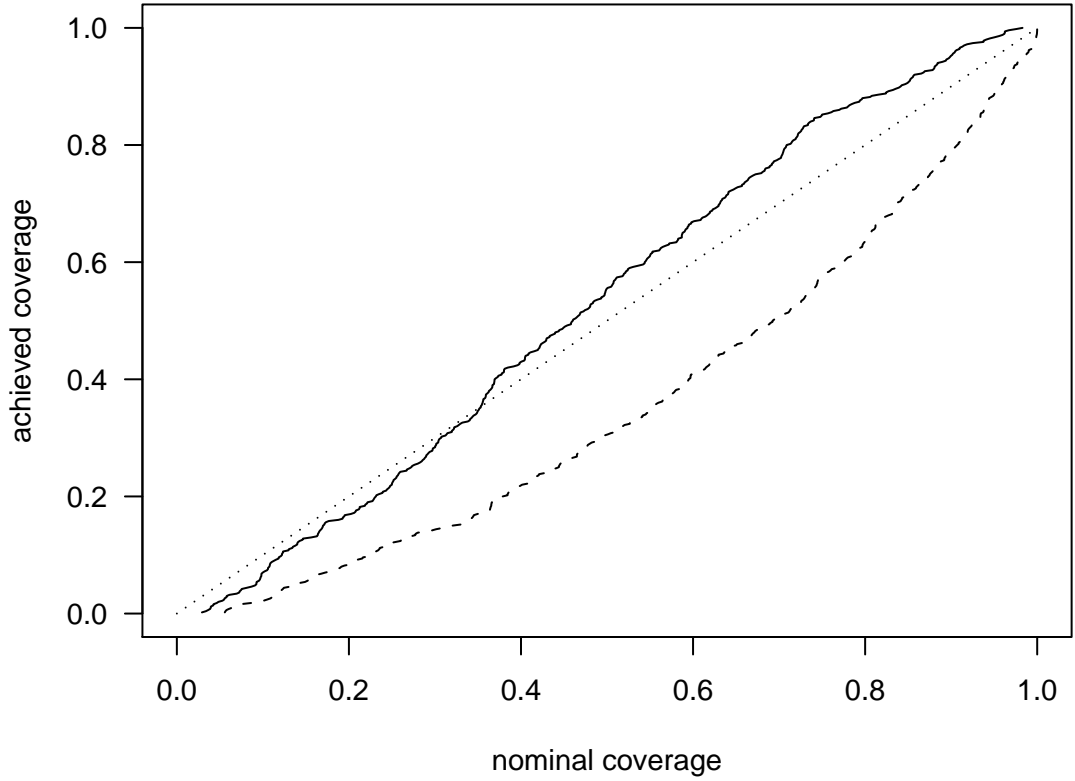}
  \end{center}
  \caption{Nominal and achieved coverage of credibility intervals for
    position of QTL, when there are errors in allele frequency
    estimation, with $\sigma=0.2$ and $r=2$.  Credibility intervals
    were constructed either without (dotted line) or with (solid line)
    the approximate correction factor of \citet{mcpeek1999}.}
  \label{fig:ncoverage}
\end{figure}

Figures~\ref{fig:coverage} and~\ref{fig:ncoverage} show results from the 500 datasets simulated with $g=4$.  The coverage
of credibility intervals constructed from the marginal posterior for
$\mu$ falls well below nominal levels.  This suggests strongly that
the simple model used for the analyses is not a good approximation to
the more realistic model the data were simulated under.  One way to
improve the model would be to allow a more realistic model for the
shape of the genealogy at the QTL.  To explicitly model this genealogy
and hence the joint distribution of breakpoints between ancestral and
nonancestral chromosome would require something like the MCMC sampler
of \citet{rannalareeve2001} or \citet{morris2002}.  Although taking
such an approach is highly desirable, it may not scale well to large
datasets and it seems worthwhile to investigate approximations.

One approximation is the ``pairwise correction'' derived by
\citet{mcpeek1999} and justified by them by the use of a quasi-score
function, and used in a Bayesian context by \citet{morris2000}.
Essentially, this involves flattening the likelihood function by
raising all likelihoods to a power $w_n=(1+(n-1)c_n)^{-1}<1$.  Here
$c_n$ is the pairwise correlation over sampled chromosomes of the
conditional score function for the position of the QTL.  An expression
for $c_n$ for a coalescent model is given in appendix D of
\citet{mcpeek1999}.  The $n$ in this equation (which $c_n$ also
depends on) is the number of chromosomes carrying the QTL.  It is not
at all clear whether or how this correction should be applied in the
present context, because (i) as noted by \citet{morris2000} the
quasi-score justification used by \citet{mcpeek1999} does not apply in
a Bayesian setting, (ii) in the present work the likelihood is never
written as a conditional product across chromosomes carrying the QTL,
(iii) it is not known how many chromosomes carry the QTL, and (iv) in
my computational implementation no proper likelihoods are ever
calculated, only likelihoods marginal to $(\pi_1,\ldots,\pi_L)$.
However, the following ad-hoc approach does produce corrected
credibility intervals that achieve coverage very close to their
nominal levels.  The procedure is to first estimate $n$ by
$n_d\mathrm{E}{(\rho)}$, the product of the number of case chromosomes
and the posterior expectation of $\rho$, and then to flatten the
marginal posterior for $\mu$ by raising it to a power $w_n$ and
renormalising.  For the simulations performed here, $w_n$ had median
$0.56$ and interquartile range 0.48--0.63.  When this procedure was
applied, good agreemement between nominal and achieved coverage is
obtained (figures~\ref{fig:coverage} and \ref{fig:ncoverage}).  This
suggests that the most serious misspecification of the current model
is the assumption of a star shaped genealogy, rather than the
assumption of linkage equilibrium in nonancestral blocks or the
absence of the disease variant in the control pool.

\subsection{Application to real data}
\label{sec:appl-real-data}
It is not really possible to examine the effectiveness of the Bayesian
method described here on real data, due to a lack of relevant
published datasets.  Primarily for the purpose of comparison with
other fine scale mapping methods, I have applied it to the dataset of
\citet{hosking2002}, and to quasi-synthetic datasets generated from
that dataset.  \citet{hosking2002} collected data using individual
genotyping.  In order to pretend that the data were acquired using DNA
pooling, I use a hypergeometric error model to relate the observed
counts with missing data to the underlying full data that were not
observed.  This assumes the data are missing at random within and
across SNPs.

\begin{figure}[p]
  \begin{center}
    \includegraphics{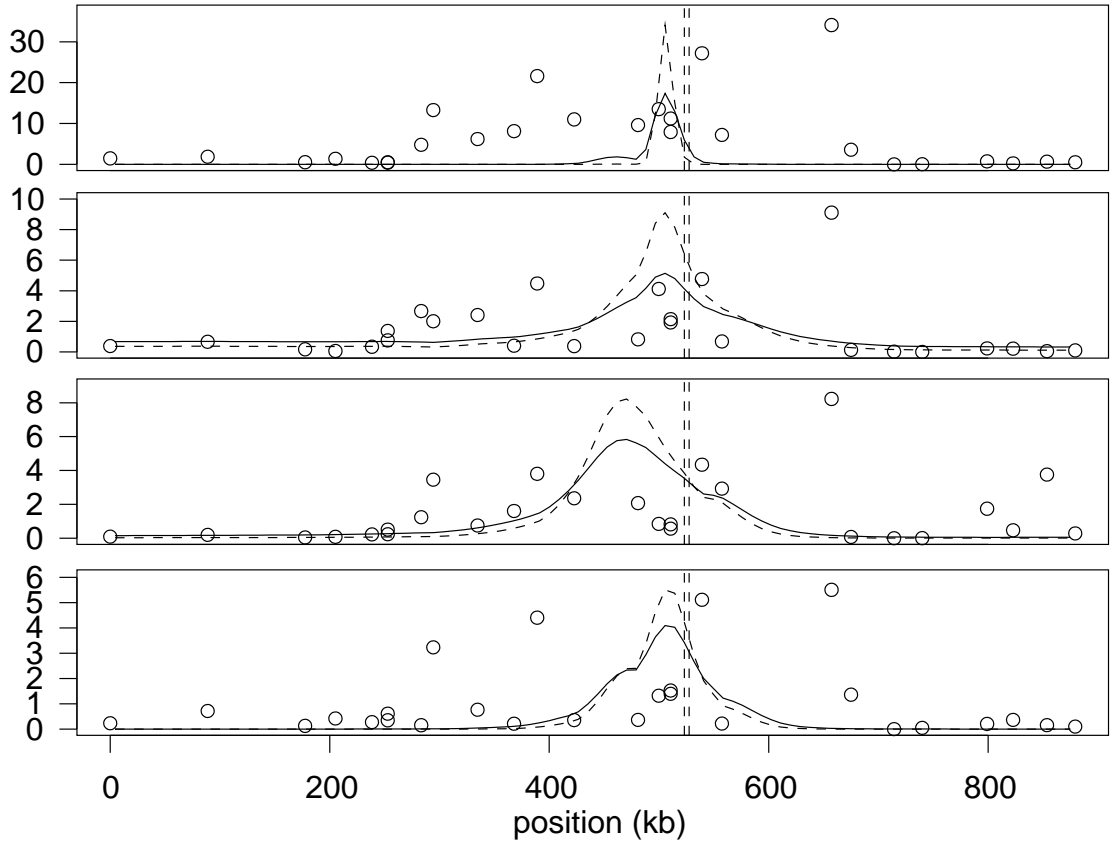}
  \end{center}
  \caption{Analysis of data of \citet[; top panel]{hosking2002}, and
    quasi-synthetic datasets generated by randomly relabelling
    controls as cases with probability 10\%, 20\% or 30\% (lower three
    panels, top to bottom). Points are $-\log_{10}{(p)}$ from single
    point $\chi^2$ tests, and dashed and solid lines are the marginal
    posterior for disease gene position, without and with the
    correction factor of \citep{mcpeek1999}.  Vertical dashed lines
    show the true position of CYP2D6.}
  \label{fig:hosking}
\end{figure}

To my knowledge, no fine scale mapping method has been published that
does not perform well on the data of \cite{hosking2002}.  Therefore,
observing that the present method performs acceptably, as shown in
figure~\ref{fig:hosking}, is not necessarily encouraging.  To simulate a disease with a complex
genetic basis, I generated three datasets by randomly relabelling
controls as cases with probability 10\%, 20\% or 30\%.  As shown in
figure~\ref{fig:hosking}, on all four datasets 95\% credibility
intervals covered the true location of the CYP2D6 gene after the
correction factor of \citep{mcpeek1999} had been applied to flatten
the posterior. This provides weak evidence that the method developed
here may be reliable for mapping QTLs from real data.

\section{Discussion}
\label{sec:discussion}
In this paper I have described and tested a Bayesian method for
detecting and mapping a QTL, using multilocus data collected using DNA
pooling within two trait groups.

Relatively recently, likelihood based fine scale mapping methods have
been developed for genotype data that build on previous haplotype
based analyses by treating the unobserved haplotypes as missing data
and integrating over all possible haplotypes that are consistent with
the observed genotypes.  This integration can be performed either
using Markov chain Monte Carlo (MCMC)
\citep{liu2001,reeverannala2002,morris2003} or using exact numerical
methods for hidden Markov models \citep{zhangzhao2002}.  Data from DNA
pools are estimated counts of alleles at each locus with no phase
information.  Fine scale mapping from genotype data and from DNA pools
can in theory be regarded as closely related missing data problems.

The approach taken in this paper combines elements of the approaches
of \citet{zhangzhao2002} and of \citet{morris2000} and
\citet{liu2001}.  Like \citet{zhangzhao2002}, I use a model that is
sufficiently simple that I can use hidden Markov model (HMM) methods
to sum over all possible haplotypes that are consistent with the
observed data.  However, after computing the likelihood using a
propagation algorithm, \citet{zhangzhao2002} then maximise that
likelihood with respect to the remaining model parameters.  In
contrast and like \citet{morris2000} and \citet{liu2001}, I embed the
HMM within a fully Bayesian approach and compute posterior probability
distributions for the quantities of interest.

One advantage of a Bayesian approach is that probability statements
can be made directly about quantities of interest.  For example, we
can state the probability that there is QTL in any given region,
including the whole region under study.  Thus, mapping and detecting a
QTL are intimately related aspects of the same analysis.  They are
different inferences that are made from the same posterior probability
distribution.  Within the Bayesian framework there is no need to
choose between a bewildering array of estimators, test statistics and
methods for correcting for multiple testing; the approach has a
pleasing simplicity, at least conceptually.

However, the probabilities computed in a Bayesian analysis are only
meaningful if the model and prior are realistic.  The Catch-22 is that
in order to compute Baysian posterior probabilities, I had to assume a
model that was worringly oversimplified and not very believable.  The
present work is therefore best regarded as a step towards Bayesian
analysis of data collected using DNA pooling.  It may be helpful to
draw parallels with methods for analysis of genotype data (collected
using individual typing).  Sadly, the present method allows us to make
inferences assuming a model less elaborate than the one of
\citet{morris2000}, whereas we might aspire to being able to assume a
model like the one of of \citet{morris2002} or
\citet{zollnerpritchard2005}.  However, Bayesian analysis of such
realistic models has required Markov chain Monte Carlo (MCMC) to
integrate over high dimensional spaces of auxiliary variables or
missing data.  Such computationally intensive approaches may have
difficulty handling large datasets.  In contrast, the method described
here is relatively fast, and large datasets could be analysed with
realistic computational resources.  For example, 27 processor-days
would be required to analyse data from a whole genome scan with 100
cases, 500,000 SNPs, and evaluation of the posterior at points 50kb
apart.  In contrast, \citet{zollnerpritchard2005} estimate that their
MCMC based procedure for data from individual typings would take 85
processor-years for the same scale of analysis.  A further advantage
of avoiding Monte Carlo methods is that the large numbers of analyses
needed for a sliding window analysis, or a permutation test, can be
performed without needing human intervention to adjust mixing
parameters or monitor convergence.  Finally and perhaps most
significantly, I am able to compute a Bayes factor (BF) to compare
models in which there is, and is not, a disease QTL in the whole
region of interest.  To my knowledge, no association mapping method
using genotype data is able to do this, although \citet{patterson2004}
are able to compute a BF for \emph{admixture} mapping using
genotype data.

There is a Bayesian justification for the present method. (``This is
the best model for which a Bayesian analysis of data from DNA pools is
currently possible.'')  However, serious concerns about model
inadequacy (``Well, that model simply isn't good enough!'') mean that,
in this paper, I have mostly focussed on the classical frequentist
justification.  Using simulations assuming a more realistic model, I
have shown that the present method is uniformly superior to classical
single point methods of analysis.  Single point methods are the most
obvious way to analyse data collected using DNA pooling, although
composite likelihood methods
\citep{terwilliger1995,xiong1997,collins1998,maniatis2004,maniatis2005}
could also be used.  The simulation results demonstrate that the BF
computed using the present method makes a more powerful test for
the presence of a QTL than the minimum $p$-value from single point
tests, that the posterior density for the position of the QTL leads to
a better point estimator than the position of the marker with the
minimum $p$-value, and that well calibrated credibility intervals can
be derived from the posterior density for the position of the QTL,
after applying the correction of \citet{mcpeek1999}.

The performance of composite likelihood (CL) methods was not examined
here.  This was because no CL method has been developed that allows
errors in allele frequency estimation, and because, to my knowledge,
no CL method assumes a model that is obviously more realistic than the
model assumed by the present method.  In particular, all CL methods
implicitly assume linkage equilibrium in non-ancestral blocks of
chromosome.  In the notation of the present paper, CL methods assume
that the number of chromosomes carrying ancestral haplotype at the
$i$-th SNP, $x_i$, is conditionally independent across SNPs.  Even a
poor model that does capture some aspect of the dependence across
SNPs, such as the star shaped genealogy assumed here, seems
preferable.  To my knowledge, there is no CL method that produces well
calibrated confidence or credibility intervals.  Perhaps because of
this, \citet{maniatis2005} state that ``[t]he main objective in
positional cloning is to estimate the kb location of a causal SNP as
accurately as possible, with its support interval an important but
secondary objective.''  However, it seems to me that we should focus
on methods for computing well calibrated credibility intervals, and
ideally a well calibrated posterior density.  The acid test is to ask
whether a statistical method informs us about what is a good action or
decision to be taken subsequently.  A point estimate for QTL position,
without a reliable measure of precision, is not very helpful for
planning future experiments to further refine the position of that
QTL.

One of the more surprising results is that, in the simulations
performed here, the nonparametric likelihood ratio (NLR) test derived
from the method proposed previously by me \citep{johnson2005a} is
basically as powerful as the BF for detecting a QTL.  This is
surprising because there is no theoretical basis for the NLR test
statistic, but a strong theoretical basis for the BF test statistic.
Since the NLR can be computed much more quickly, both in absolute and
complexity terms, its performance in simulations over a wider range of
parameters will be examined in a subsequent paper.

Given that the NLR performs as well as the BF for detecting a QTL, but
that the BF is much more expensive to compute, one might reasonably
ask what are the benefits of the Bayesian method described here.
Firstly, the BF has a Bayesian interpretation, and since it can be
negative it can indicate Bayesian sense evidence in favour of there
being no QTL.  The NLR test statistic can never be negative, has no
direct Bayesian interpretation, and is not a good approximation to the
BF.  Secondly, the posterior median from the Bayesian method provides
superior point estimates under absolute error losses.  Thirdly, the
Bayesian method produces well calibrated credibility intervals, but
the profile likelihood method I proposed previously does not
\citep[see figure 4 of][]{johnson2005a}.  Finally, the unconditional
coverage frequencies of credibility intervals say nothing about the
conditional or Bayesian sense performance of a method.  For multistage
QTL mapping experiments we should probably guide our choice of where
to type further markers using the typically complex, heavy tailed and
often multimodel posterior distributions computed using the Bayesian
method described here, as exemplified in figure~\ref{fig:nexample}.
If analysing data from a whole genome scan, I would recommend a
multistage analysis that first uses the NLR statistic to identify
regions of interest, and the to use the CPQ algorithm to compute Bayes
factors and posterior distributions for QTL position within those
regions.

Given the large number of simplifications made in specifying the model
used here, one might wonder why the method works at all.  The three
most obviously inadequate approximations are the star shaped
genealogy, the absence of the disease allele in the control pool,
and the assumption of linkage equilibrium in non-ancestral
blocks of chromosome.  I will briefly discuss these inadequacies in turn.

Figures~\ref{fig:coverage} and \ref{fig:ncoverage} show that
credibility intervals only achieve prescribed coverage levels when a
correction is made for the genealogy not in fact being star shaped.
This suggests a serious inadequacy of the model.  This is further
supported by the observation of the very similar ROC curves in
figures~\ref{fig:roc-no-error} and \ref{fig:roc-with-error} for the
theoretically optimal BF (assuming a star shaped genealogy), and the
NLR statistic that has no theoretical basis \citep[but allows any
shape genealogy;][]{johnson2005a}.  Addressing this inadequacy is
likely to lead to greater power to detect a QTL, and perhaps smaller
credibility intervals of a given size.  However, it will be hard to
achieve without imposing a substantial computational burden.  In
particular, it may become difficult to compute the BF test statistic
if MCMC is used to integrate over genealogies at the QTL.

Although it is conceptually straightforward to allow blocks of
ancestral chromosome in the control pool, this would increase the
number of hidden states at each SNP from $(\nd+1)$ to
$(\nd+1)(\nc+1)$.  Since the propagation algorithm
(section~\ref{sec:hidden-markov-model}) requires time that is
quadratic in the number of hidden states, the analysis would be
intractable using the current approach.  As an alternative, any number of
separate pools could be treated as conditionally independent HMMs, but
then we would have to integrate over the high dimensional space of
allele frequencies and ancestral haplotypes using MCMC or importance
sampling (see below).

It is possible that the current model adapts to fit there being blocks
of ancestral chromosome in the control pool, by appropriate adjustment
of the allele frequency parameters.  Ancestral blocks that are
explicitly modelled in the disease pool would then represent
additional blocks beyond what would be expected according to the
adjusted allele frequency parameters.  If this was so, the parameter
$\rho$ might be best interpeted as representing the rate of
\emph{excess} disease alleles in the disease pool.

Since only marginal observations are available, the assumption of
linkage equilibrium may be relatively innocuous.  Since there is
virtually no information in the data about linkage disequilibrium,
introducing parameters describing linkage disequilibrium into the
model might have little effect on inferences about the quantities of
interest.  It is possible to retain the present framework where all
the data are modelled as a single HMM, but to include pairwise linkage
disequilibrium by allowing allelic state along each chromosome to be a
first order Markov chain \citep[see e.g.][]{liu2001,morris2002}.  This
will be quite computationally expensive, but could be examined in the
future.

For the parameters chosen for the simulations performed here, the
benefits of the present Bayesian method are somewhat modest.  It
remains unclear whether there would be larger benefits for other
values of the simulation parameters, in particular more SNPs in the
dataset, and/or larger benefits from a Baysian analysis with a more
realistic model.  Clarification of both points awaits access to
substantial computational resources.  It is worth commenting that
many of the variables in the present model also feature in more
elaborate models, and therefore the present approach could be used to
generate (for example) a joint importance sampling distribution for
the ancestral haplotype, allele frequencies, and age and position of
the QTL.

Even the simulated datasets studied here were generated under a model
that lacks realism in several respects.  For example, in simulating
errors in allele frequency estimation I have ignored differential
amplification of the two alleles, which may cause estimates of allele
frequencies obtained using DNA pools to be biased.  This manifests
itself as only a second order effect on the difference in allele
frequency between case and control pools \citep{visscher2003}.
Differential amplification can be accomodated easily in the present
method of analysis, for example by making \ey{d}{i} a vector
consisting of data from the pool and also from heterozygous
individuals or pools of known composition.  Even if no data from
heterozygotes is available, it is possible to compute a
$\Pr{(\hat{y}|y)}$ by integrating over a distribution of differential
amplification constants, like in the approach of \citet{moskvina2005}.

One feature of the posteriors calculated using the present method (and
especially after \citet{mcpeek1999} flattening) is that they are very
heavy tailed, and so large credibility intervals (99\%, 99.9\%) tend
to be very wide, perhaps almost as wide as credibility intervals
computed from the prior!  This suggests that, if a series of fine
scale mapping experiments were conducted using DNA pooling, we would
not be making radical reductions in the size of the region under study
at each stage, but rather would be increasing the density of markers
in some regions more than others after each stage of analysis.

\section*{Software}
A software package implementing the methods described here is
available from the web site
\texttt{http://homepages.ed.ac.uk/tobyj/software/}~.  Source code is
available and the package can be distributed freely under the terms of
the GNU general public licence \citep{fsf1991}.

\bibliography{tobyrefs}

\clearpage
\begin{deluxetable}{l p{0.8\textwidth}}
  \tablecaption{Frequently used notations.\label{tab:notations-used}}
  \tablehead{
    \colhead{Symbol} & \colhead{Meaning}}
  \startdata
  $a_i$ & Allele present (0 or 1) on ancestral haplotype at $i$-th SNP \\
  $b$,$b'$ & Backwards variables, see (\ref{eq:bvdefright}) and (\ref{eq:bvdefleft}) \\
  $\betadist{(\alpha,\beta)}$ & Beta distribution with parameters $\alpha$ and $\beta$ \\
  $\bindist{(n,p)}$ & Binomial distribution with parameters $n$ and $p$ \\
  $\bindens{(x,n,p)}$ & Probability of drawing $x$ from a binomial distribution with parameters $n$ and $p$ \\
  $\mbox{BF}$ & Bayes factor \\
  CPQ & Cartesian product quadrature \\
  $d_\mu$ & Number of design points used for $\mu$ in quadrature algorithm \\
  $e_i$ & Precision of assay used to genotype $i$-th SNP \\
  $\expdist{(\lambda)}$ & Exponential distribution with rate parameter $\lambda$ (mean $1/\lambda$) \\
  $g$ & Genotype relative risk; factor by which disease allele
  increases penetrance or risk \\
  $\gamdist{(\alpha,\beta)}$ & Gamma distribution with shape parameter $\alpha$ and scale parameter $\beta$ \\
  HMM & Hidden Markov model \\
  $i$ & Index of SNP, $i=1,\ldots,L$ \\
  i.b.d. & Identical by descent \\
  $L$ & Number of SNPs \\
  $m_i$ & Map position of the $i$-th marker \\
  MCMC & Markov chain Monte Carlo \\
  $\nc$, $\nd$ & Number of chromosomes in control and case pools respectively \\
  $\ndist{(\mu,\sigma^2)}$ & Normal distribution with mean $\mu$ and variance $\sigma^2$ \\
  $\mbox{NLR}$ & Nonparametric likelihood ratio, see (\ref{eq:proftestdef}) \\
  $\pmin$ & Smallest $p$-values out of $L$ tests in single point analysis\\
  $P_{i,a}$ & Prior parameter: $\p{i}{a}\sim\betadist{(P_{i,a},P_{i,1-a})}$ \\
  $r$ & Number of experimental replicates used to estimate $\Delta\widehat{C_t}$\\
  $R$ & Prior parameter: $\rho\sim\betadist{(R_1,R_0)}$ \\
  ROC & Receiver operating characteristics \\
  $T$ & Prior parameter: $\tau\sim\expdist{(T)}$ \\
  $x_i$ & Number of chromosomes in case pool carrying ancestral
  i.b.d.\ haplotype at $i$-th SNP \\
  $x_\mu$ & Number of chromosomes in case pool carrying ancestral i.b.d.\ haplotype at position of QTL \\
  \ya{c}{i}{a} & True count of allele $a$ at $i$-th SNP in control pool \\
  \ya{d}{i}{a} & True count of allele $a$ at $i$-th SNP in case pool \\
  \eya{c}{i}{a} & Estimated count of allele $a$ at $i$-th SNP in control pool \\
  \eya{d}{i}{a} & Estimated count of allele $a$ at $i$-th SNP in case pool \\
  $\hat{y}$ & Shorthand for \eya{c}{i}{1} or \eya{d}{i}{1} for some $i$ \\
  $\hat{\vec{y}}$ & All the data \\
  $\alpha$ & Nominal size (rate of type I error) of a test \\
  $\Delta\widehat{C_t}$ & Estimated lag between PCR growth curves used to type DNA pool\\
  $\gamma_G$ & Penetrance (risk of disease) for genotype $G$ at the QTL\\
  $\mu$ & Map position of the disease locus \\
  $\muminp$ & Map position of SNP with smallest $p$-value in single
  point analysis \\
  $\p{i}{1}$ & Expected frequency of allele 1 at $i$-th SNP in
  non-ancestral chromosome \\
  $\rho$ & Expected frequency of disease allele in case pool \\
  $\sigma$ & Standard deviation of experimental error in estimation of
  $\Delta\widehat{C_t}$ \\
  $\tau$ & Age of the disease allele \\
  \enddata
\end{deluxetable}

\begin{deluxetable}{l l l l l l l l}
  \tablecaption{Performance of tests to detect a disease QTL, when allele
    frequencies in each pool are known exactly.
    \label{tab:power-no-error}}
  \tablehead{
    \colhead{Statistic} & \colhead{Method} & \colhead{Nominal size} &
    \colhead{Critical value} & \colhead{True size} & 
    \colhead{Power} }
  \startdata
  $2\ln{\mbox{BF}}$&Arbitrary&&0&0.080&0.870\\
  &&&&(0.058, 0.107)&(0.837, 0.898)\\
  $2\ln{\mbox{BF}}$&Arbitrary&0.05\tablenotemark{a}&5.889&0.010&0.710\\
  &&&&(0.003, 0.023)&(0.668, 0.749)\\
  $p_{\min}\times L$&Bonferonni&0.05&0.05&0.040&0.720\\
  &&&&(0.025, 0.061)&(0.678, 0.759)\\
  $2\ln{\mbox{BF}}$&Simulation&0.05&0.903&0.050&0.842\\
  &&&&(0.033, 0.073)&(0.807, 0.873)\\
  $p_{\min}\times L$&Simulation&0.05&0.063&0.050&0.740\\
  &&&&(0.033, 0.073)&(0.699, 0.778)\\
  $2\ln{\mbox{BF}}$&Arbitrary&0.01\tablenotemark{a}&9.19&0.002&0.628\\
  &&&&(0.000, 0.011)&(0.584, 0.67)\\
  $p_{\min}\times L$&Bonferonni&0.01&0.010&0.000&0.582\\
  &&&&(0.000, 0.006)&(0.537, 0.626)\\
  $2\ln{\mbox{BF}}$&Simulation&0.01&5.864&0.010&0.710\\
  &&&&(0.003, 0.023)&(0.668, 0.749)\\
  $p_{\min}\times L$&Simulation&0.01&0.027&0.010&0.664\\
  &&&&(0.003, 0.023)&(0.621, 0.705)\\
  \enddata
  \tablenotetext{a}{not a nominal size in the classical sense but a nominal upper
    bound on the Bayesian sense error rate}
\end{deluxetable}

\begin{deluxetable}{l l l l l l l l}
  \tablecaption{Performance of tests to detect a disease QTL, when there
    are errors in allele frequency estimation with $r=2$ replicates and
    $\sigma=0.2\mbox{ PCR cycles}$.
    \label{tab:power-error}}
  \tablehead{
    \colhead{Statistic} & \colhead{Method} & \colhead{Nominal size} &
    \colhead{Critical value} & \colhead{True size} & 
    \colhead{Power} }
  \startdata
  $2\ln{\mbox{BF}}$&Arbitrary&&0&0.080&0.782\\
  &&&&(0.058, 0.107)&(0.743, 0.817)\\
  $2\ln{\mbox{BF}}$&Arbitrary&0.05\tablenotemark{a}&5.889&0.008&0.532\\
  &&&&(0.002, 0.020)&(0.487, 0.576)\\
  $p_{\min}\times L$&Bonferonni&0.05&0.05&0.040&0.560\\
  &&&&(0.025, 0.061)&(0.515, 0.604)\\
  $2\ln{\mbox{BF}}$&Simulation&0.05&0.723&0.050&0.746\\
  &&&&(0.033, 0.073)&(0.705, 0.784)\\
  $p_{\min}\times L$&Simulation&0.05&0.085&0.050&0.642\\
  &&&&(0.033, 0.073)&(0.598, 0.684)\\
  $2\ln{\mbox{BF}}$&Arbitrary&0.01\tablenotemark{a}&9.19&0.000&0.424\\
  &&&&(0.000, 0.006)&(0.380, 0.469)\\
  $p_{\min}\times L$&Bonferonni&0.01&0.01&0.010&0.432\\
  &&&&(0.003, 0.023)&(0.388, 0.477)\\
  $2\ln{\mbox{BF}}$&Simulation&0.01&4.28&0.010&0.596\\
  &&&&(0.003, 0.023)&(0.552, 0.639)\\
  $p_{\min}\times L$&Simulation&0.01&0.01&0.010&0.432\\
  &&&&(0.003, 0.023)&(0.388, 0.477)\\
  \enddata
  \tablenotetext{a}{not a nominal size in the classical sense but a nominal upper
    bound on the Bayesian sense error rate}
\end{deluxetable}

\begin{deluxetable}{r r r r}
  \tablecaption{Performance of point estimators of QTL position.\label{tab:point-exp-loss}}
  \tablehead{
    Estimator&\multicolumn{2}{c}{Average
      expected loss under}\\\cline{2-3}
    &squared error losses&absolute error losses}
  \startdata
  \cutinhead{Allele frequencies known exactly}
  \muminp&0.208$\,{}^2$&0.120\\
  $\expn{(\mu|\hat{\vec{y}})}$&0.165$\,{}^2$&0.107\\
  $\medn{(\mu|\hat{\vec{y}})}$&0.166$\,{}^2$&0.105\\
  NP method&0.165$\,{}^2$&0.112\\
  \cutinhead{Errors in allele frequency estimation}
  \muminp&0.239$\,{}^2$&0.146\\
  $\expn{(\mu|\hat{\vec{y}})}$&0.195$\,{}^2$&0.132\\
  $\medn{(\mu|\hat{\vec{y}})}$&0.203$\,{}^2$&0.130\\
  NP method&0.198$\,{}^2$&0.136\\
  \enddata
\end{deluxetable}

\clearpage
\listoffigures

\end{document}